\documentclass[a4paper,fleqn,usenatbib]{mnras}

\usepackage{newtxtext,newtxmath}
\usepackage[T1]{fontenc}
\usepackage{ae,aecompl}

\usepackage{graphicx}	% Including figure files
\usepackage{amsmath}	% Advanced maths commands
\title[Estimating age-metallicity distributions]{Estimating the age-metallicity distribution of a stellar sample from the probability distributions of the individual stars}

\author[C. L. Sahlholdt \& L. Lindegren]{
Christian L. Sahlholdt$^{1}$\thanks{E-mail: sahlholdt@astro.lu.se}
and Lennart Lindegren$^{1}$
\\
$^{1}$Lund Observatory, Department of Astronomy and Theoretical Physics, Box 43,
      SE-221 00 Lund, Sweden\\
}

\date{Accepted XXX. Received YYY; in original form ZZZ}

\pubyear{2020}

\begin{document}
\label{firstpage}
\pagerange{\pageref{firstpage}--\pageref{lastpage}}
\maketitle

\begin{abstract}
Estimating age distributions, or star formation histories, of stellar populations in the Milky Way is important in order to study the evolution of trends in elemental abundances and kinematics.
We build on previous work to develop an algorithm for estimating the age-metallicity distribution which uses the full age-metallicity probability density functions (PDFs) of individual stars.
No assumptions are made about the shape of the underlying distribution, and the only free parameter of the algorithm is used to ensure a smooth solution.
In this work we use individual age-metallicity PDFs from isochrone fitting of stars with known metallicities.
The method is tested with synthetic samples and is found to recover the input age-metallicity distribution more accurately than the distribution of individually estimated ages and metallicities.
The recovered sample age distribution is always more accurate than the distribution of individual ages, also when restricted to the most precise individual ages.
By applying the method to the stars in the Geneva-Copenhagen survey, we detect a possible minimum in the star formation history of the Solar neighbourhood at an age of 10~Gyr which is not seen in the distribution of individual ages.
Although we apply the method only to age-metallicity distributions, the algorithm is described more generally and can in principle be applied in other parameter spaces.
It is also not restricted to individual parameter distributions from isochrone fitting, meaning that a sample age distribution can be estimated based on individual age PDFs from other methods such as asteroseismology or gyrochronology.
\end{abstract}


%%%%%%%%%%%%%%%%%%%%%%%%%%%%%%%%%%%%%%%%%%%%%%%%%%

%%%%%%%%%%%%%%%%% BODY OF PAPER %%%%%%%%%%%%%%%%%%

\section{Introduction}
Our knowledge about the structure and evolution of the Milky Way is quickly improving thanks to a large number of both photometric and spectroscopic stellar surveys.
Ground-based spectroscopic surveys provide elemental abundances while the \textit{Gaia} mission \citep{2016A&A...595A...1G} provides distances and kinematic properties through precise astrometric observations.
When combined, these data give insight into the overall structure of the stellar content of the Galaxy, and put constraints on the star formation history (SFH) \citep[e.g.,][]{2019A&A...624A..19B}.
The survey data has also aided the development and interpretation of cosmological galaxy models.
Such development has driven considerable progress in the reconstruction of the SFH of the Milky Way and of its modelling in a cosmological context both by means of semi-analytic models and hydro-simulations \citep[e.g.,][]{2009MNRAS.400.1347C,2012MNRAS.422..215Y,2012MNRAS.427.1401C,2014MNRAS.444.3845F,2020MNRAS.491.5435B}.
The most recent hydrodynamic simulations of Milky Way-like galaxies explain the observed bimodality in the $[\alpha/\mathrm{Fe}]$ versus $[\mathrm{Fe}/\mathrm{H}]$ abundance plane as a combination of radial variations in the SFH of the Galactic disc, accretion of metal-poor gas, and migration of stars \citep{2020MNRAS.496...80V,2020arXiv200606008A}.
These new insights are significantly advancing our understanding of the evolution of the Milky Way.

Information about the Galaxy's evolution can also be gained by estimating the SFH of an observed stellar population.
Colour-magnitude diagram (CMD) fitting is one method for estimating the SFH of a sample of stars with known absolute magnitudes and colors \citep[e.g.,][]{2002MNRAS.332...91D, 2005ARA&A..43..387G}.
This method is based on analysis of the density of stars across the CMD.
A large number of synthetic stars with a uniform distribution in age and metallicity are generated from isochrones.
These stars make up a model CMD which is divided into a number of partial models with narrow intervals in age and metallicity.
For each partial model, the number of stars in each colour-magnitude bin across the CMD is computed.
The SFH can then be estimated by finding the linear combination of partial models in each CMD bin that best matches the observed number of stars.
CMD fitting has been used to study the SFH of dwarf galaxies with resolved stellar populations \citep{2009ARA&A..47..371T}.
It has also recently been applied to the old Milky Way population using CMDs based on \textit{Gaia} photometry and parallaxes \citep{2019NatAs...3..932G}.

When metallicities are available for the sample, in addition to colours and magnitudes, one can estimate individual ages instead and estimate the SFH using their distribution.
These age estimates are usually made with Bayesian isochrone fitting \citep{2005A&A...436..127J}.
For each star, a probability density function (PDF) over the age is computed, and a single age estimate is obtained by taking e.g. the mode or mean (expectation value) of the distribution.
With individual isochrone fitting it is also possible to include asteroseismic parameters which put stronger constraints on the mass and therefore also the age \citep{2013ARA&A..51..353C}.
This is an advantage over CMD fitting which only utilises the age information contained in the colours and magnitudes.
A disadvantage of individual ages from isochrone fitting is that they can be very uncertain for stars that are not near the turnoff in the CMD.
This is because isochrones of different ages overlap significantly on the lower main sequence and on the giant branch leading to very broad, sometimes even flat, age PDFs.

Even with large samples of individual ages, the SFH will be smeared out by the most uncertain ages.
This can be mitigated to some degree by only choosing the stars with the most precise ages; however, this introduces a selection effect favouring stars in certain regions of the CMD, and favouring young stars over older ones.
This motivates the use of the full age PDF for each star to inform the SFH.
Such a method was described by \citet{1999MNRAS.304..705H}.
With this method, the SFH which best explains the complete set of individual age PDFs is found without any prior assumptions about its form.
\citet{2005ESASP.576..171J} demonstrated how this method can distinguish between different SFHs which look more or less identical in the distributions of individual ages.

Since then, several other methods using the full age PDFs from isochrone fitting have been presented.
\citet{2013MNRAS.435.2171W} used Bayesian inference of the SFH parametrized by an unknown number of Gaussian bursts.
With this method, which is aimed at and tested for young populations (<500 Myr), each star can be assigned a probability of belonging to each of these bursts.
\citet{2016ApJ...817...40F} likewise used Bayesian inference of the SFH history but parametrized it as a Gaussian for different bins in alpha element abundance.
This is useful for quantifying the relation between age and abundances, but it does not capture deviations of the SFH from a Gaussian.
A novel method by \citet{2019A&A...629A.127M} represents the sample age PDF, obtained by summing all individual age PDFs, by a linear combination of age PDFs of synthetic mono-age populations.
The coefficients of the linear combination define the SFH.
Mono-age populations are defined to follow a similar parameter distribution as the real sample.
The advantage of this method is that it removes some of the biases associated with isochrone fitting; however, it requires one to construct mono-age populations for each sample that is analysed.
\citet{2013MNRAS.428..763S} expand the idea further by including the metallicity as a second dimension.
That is, they define the probability of a single star belonging to a certain age-metallicity distribution as a linear combination of probabilities of belonging to different isochrones.
The age-metallicity distribution is estimated by optimizing the coefficients in the linear combination for the entire sample.

In this work we build on the methods developed by \citet{1999MNRAS.304..705H} and \citet{2005ESASP.576..171J} and present an inversion algorithm for estimating the sample parameter PDF which works in multiple dimensions.
We apply it to the specific (two-dimensional) case of estimating the age-metallicity distribution of a population of stars, without any prior assumptions about its form, based on the individual age-metallicity PDFs.
The individual PDFs used for testing are obtained by isochrone fitting to parameters available from large spectroscopic surveys combined with \textit{Gaia} astrometry: $T_{\mathrm{eff}}$, [Fe/H], apparent magnitude, and parallax.
Since the metallicity is included in the fit, we are essentially deriving the SFH in metallicity bins.
However, the age-metallicity distribution is obtained directly, with metallicity uncertainties taken into account, and the metallicity bins are defined by the resolution of the individual PDFs.
Additionally, the inversion algorithm is described independently from the method used to obtain the individual PDFs, and without any constraints on the number of dimensions.
This opens up the possibility of adding more parameters in the future (e.g. expanding it to three dimensions and age-metallicity-alpha distributions), and of combining age PDFs (in the one-dimensional case) derived with different methods into a single SFH.

The paper is organized as follows.
In section~\ref{sec:method} the methods for estimating individual and sample age-metallicity distributions are introduced.
The method is tested with synthetic samples in section~\ref{sec:synth_intro}, starting with samples drawn from single isochrones before testing the recovery of extended age-metallicity distributions.
In the final subsection, we compare the recovered sample SFHs with the distributions of individual age estimates.
We test the method with a real data set in section~\ref{sec:real_data_test} by applying it to a sample of nearby stars in the Geneva-Copenhagen survey.
In section~\ref{sec:discussion} we discuss the performance of the method and how to apply it in practice before giving our conclusions in section~\ref{sec:conclusion}.

\section{Method} \label{sec:method}
The method is divided into two steps described in the following subsections.
First, a grid of stellar isochrones is fitted to the observables of each star in order to obtain the individual age-metallicity PDFs.
Then the individual PDFs are combined to estimate the sample distribution.
The second step is described in general terms because, in principle, the method can be applied to distributions over any parameters with any number of dimensions, and the individual distributions can be derived by any means.
In this work, however, we only implement and test it in the two-dimensional case of age-metallicity distributions estimated by isochrone fitting.

\begin{figure}
\centering
\includegraphics[width=\columnwidth]{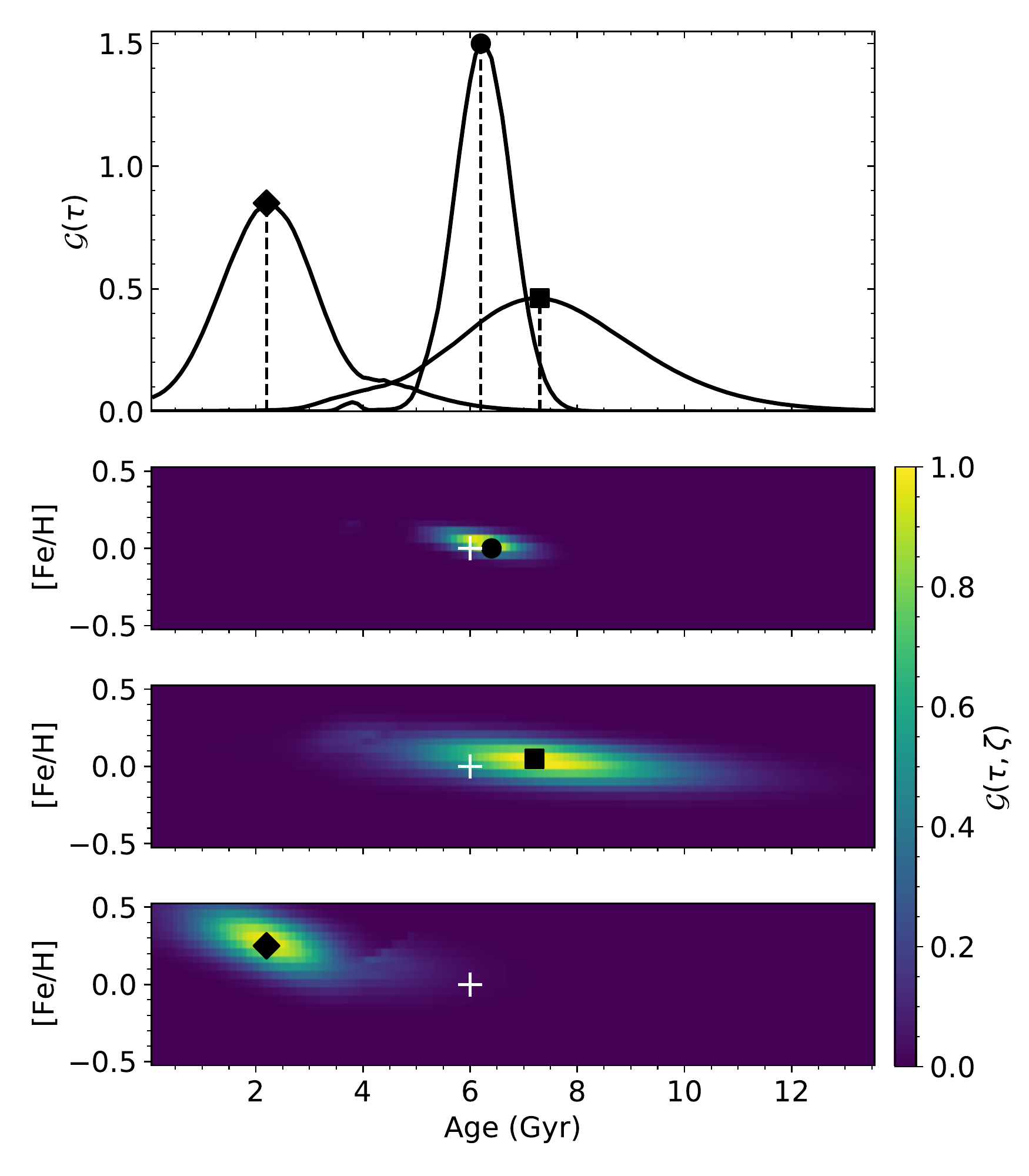}
\caption{Three examples of 2D (age-metallicity) $\mathcal{G}$~functions and the corresponding 1D (age) $\mathcal{G}$~functions obtained by marginalizing the metallicity with a flat prior.
These examples are based on isochrone fitting of the same synthetic star with three different levels of parameter uncertainties (see \autoref{fig:synth_HR} for its position in the HR~diagram).
The true age and metallicity (6~Gyr and 0~dex) are marked by the white cross.
In each example the black marker indicates the mode of the distribution in both the 2D and 1D case, and the black dashed lines indicate the individual age estimates.}
\label{fig:gfuncs_ex}
\end{figure}

\subsection{Individual probability densities}
The method used to estimate individual age-metallicity PDFs has already been described in detail by \citet[][see their Appendix A for the algorithm]{2019A&A...622A..27H}.
This method is based on Bayesian isochrone fitting and gives the joint marginal likelihood $\mathcal{G}(\tau, \zeta | \mathbf{x})$ of age, $\tau$, and metallicity, $\zeta$\footnote{In this work the metallicity in the stellar models is defined as $\zeta = \log(Z/X)-\log(Z/X)_{\odot}$.}, given a set of observables $\mathbf{x}$.
In short, we use a grid of isochrones where each stellar model is described by its initial mass, $m$, age, $\tau$, metallicity, $\zeta$, and distance modulus, $\mu$.
For each model, the likelihood given a set of observables, $\mathbf{x}$, is calculated as
\begin{equation}
    L(m, \tau, \zeta, \mu | \mathbf{x}) = \exp\left[ -\dfrac{1}{2} \sum_{i} \left( \dfrac{x_i-X_i(m, \tau, \zeta, \mu)}{\sigma_i} \right)^2 \right] \, ,
\end{equation}
where $X_i$ is the model prediction of the observable $x_i$ with uncertainty $\sigma_i$.
$\mathcal{G}(\tau, \zeta | \mathbf{x})$ is obtained by marginalizing over the mass and distance modulus with priors $\xi(m)$ and $\psi(\mu)$
\begin{equation}
    \mathcal{G}(\tau, \zeta | \mathbf{x}) \propto \int_{\mu}\psi(\mu)d\mu \int_{m}\xi(m)L(m, \tau, \zeta, \mu | \mathbf{x})dm \, .
\end{equation}
In this work a Salpeter IMF is used as the mass prior, and the distance modulus prior is flat.
The one-dimensional $\mathcal{G}(\tau | \mathbf{x})$ is calculated by marginalizing over the metallicity with a flat prior.
These $\mathcal{G}$~functions have the property that the product $\mathcal{G}(\tau, \zeta | \mathbf{x})\phi(\tau, \zeta)$ is proportional to the joint posterior PDF for $\tau, \zeta$, if $\phi$ is their prior density.

\autoref{fig:gfuncs_ex} shows three examples of $\mathcal{G}(\tau, \zeta)$ for a synthetic star drawn from an isochrone with an age of 6~Gyr and a metallicity of 0~dex.
We refer to the age and metallicity of the isochrone from which the star was drawn as the star's true parameters.
The difference between the examples is that the star has had its true parameters perturbed by different levels of uncertainties (as explained in \autoref{sec:synth_intro}) before calculating $\mathcal{G}(\tau, \zeta)$.
To get an estimate of the star's age, we can take the mode of $\mathcal{G}(\tau | \mathbf{x})$ as shown in the upper panel of \autoref{fig:gfuncs_ex}.
At the lowest level of uncertainty (circle) the age estimate is more precise (the $\mathcal{G}$~function is narrower) and accurate (the mode is closer to the true age) than at higher uncertainties (square and diamond).
Any age estimate obtained from a single $\mathcal{G}$~function will be referred to as an individually estimated age.
Such an estimate has an associated uncertainty, $\sigma_{\tau}$, which we estimate as half of the 68 per cent confidence interval defined following \citet{2005A&A...436..127J}.
This confidence interval is the region within which the $\mathcal{G}$~function is greater than 0.61 times the mode, corresponding to $\pm 1 \sigma$ for a Gaussian distribution.

The $\mathcal{G}$~function can of course be defined more generally as a function of any set of parameters, $\theta$, and this is the notation we will use in the following general description of the method.
In this work we will only consider the two-dimensional case where $\theta = (\tau, \zeta)$, i.e. joint age-metallicity distributions.

\subsection{Sample parameter distribution} \label{sec:method_sample_dist}
We now consider a sample of $n$ stars assumed to be independently drawn from a population where the set of parameters $\theta$ has a certain unknown PDF $\phi(\theta)$.
The goal is to estimate $\phi(\theta)$ given the individual $\mathcal{G}$~functions, $\mathcal{G}_{i}(\theta)$, $i=1\dots n$.
In this notation, the dependence on the data $\mathbf{x}$ is suppressed for brevity.
Each $\mathcal{G}_{i}$ has an arbitrary normalization; however, $\phi$ is a proper density and must satisfy
\begin{equation} \label{eq:norm_req}
    \int\phi(\theta)\mathrm{d}\theta = 1 \, .
\end{equation}

Assuming the prior density $\phi(\theta)$, the posterior PDF of $\theta$ for a single star $i$ is
\begin{equation}
    g_{i}(\theta) = \frac{\mathcal{G}_{i}(\theta)\phi(\theta)}{\int \mathcal{G}_{i}(\theta)\phi(\theta)\mathrm{d}\theta} \, .
\end{equation}
In the Bayesian context, the denominator in this expression,
\begin{equation}
    L_{i}(\phi) = \int \mathcal{G}_{i}(\theta)\phi(\theta)\mathrm{d}\theta \, ,
\end{equation}
is called the evidence, because in hypothesis testing it can be used to compare how well different models agree with the data.
It is possible to interpret the evidence, $L_{i}(\phi)$, as the likelihood of $\phi$ given the data, $\mathcal{G}_{i}(\theta)$, of the $i$-th star.
For the whole sample of $n$ stars, the total likelihood is then
\begin{equation} \label{eq:phi_likelihood}
    L(\phi) = \prod_{i} \int \mathcal{G}_{i}(\theta)\phi(\theta)\mathrm{d}\theta \, .
\end{equation}
This is the equation used by e.g. \citet{1999MNRAS.304..705H} in the one-dimensional case with $\theta = \tau$.
The problem is now to find the non-negative function $\hat{\phi}(\theta)$ which minimizes
\begin{equation} \label{eq:lnL}
    -\ln L = -\sum_{i} \ln\left( \int \mathcal{G}_{i}(\theta)\phi(\theta)\mathrm{d}\theta \right) \, ,
\end{equation}
subject to the constraint of equation \eqref{eq:norm_req}.

This is an inversion problem and it does not have a unique solution since we are attempting to recover an unknown continuous function from a finite set of measurements.
Therefore, we add a regularization term, which can be varied in strength, to favor smoother solutions.
We add a term like the one used by \citet{2005ESASP.576..171J}, which penalizes large second derivatives of $\phi$, only here it is generalised to any number of dimensions.
The problem is then to minimize the regularized expression
\begin{align} \label{eq:lnL_final}
        -\ln L' = - \sum_{i} \ln\left( \int \mathcal{G}_{i}(\theta)\phi(\theta)\mathrm{d}\theta \right) + \alpha \int\left(\sum_{q=1}^{N_{\mathrm{dim}}}s_{q}^{2}\frac{\partial^2\phi}{\partial\theta_{q}^{2}} \right)^2\mathrm{d}\theta \, ,
\end{align}
where $\theta_{q}$, $q=1\dots N_{\mathrm{dim}}$, are the individual parameters and $s_{q}$ is a suitable scaling factor with the same dimensions as $\theta_{q}$.

For a given value of $\alpha$, we minimize equation \eqref{eq:lnL_final} by finding the root of the derivative with respect to $\phi$ using Newton-Raphson iteration.
The simple Newton-Raphson algorithm will only converge to the global minimum if the initial approximation is sufficiently close to the solution.
Therefore, we start by forcing the solution to be a constant function and slowly relax this requirement until convergence.
A detailed description of this implementation, including the discretization of equation \eqref{eq:lnL_final}, is given in Appendix~\ref{sec:appA}.

The optimal choice of $\alpha$ is not always straightforward, and it varies depending on the problem at hand.
This choice is discussed separately in Appendix \ref{sec:appB}.
For all examples shown in this work, $\alpha$ has been chosen as laid out in that section unless otherwise stated.

In the specific application considered in this work, with $\theta = (\tau, \zeta)$, we refer to the solution $\hat{\phi}$ as the sample age-metallicity distribution (SAMD).
The one-dimensional sample age distribution (SAD), which gives an estimate of the SFH, can be obtained by marginalizing the metallicity in the SAMD, i.e. by summing over the metallicity at each age, or directly by applying the method with $\theta = \tau$.

\section{Synthetic data tests} \label{sec:synth_intro}
In this section the algorithm is tested on different sets of synthetic stellar samples drawn from a grid of PARSEC isochrones \citep{2012MNRAS.427..127B}.\footnote{Downloaded from \url{http://stev.oapd.inaf.it/cgi-bin/cmd}}
Naturally, this same isochrone grid is used in the fitting step to obtain individual $\mathcal{G}$~functions of the synthetic stars.
For each star in the sample, we define the values of the observables $\mathbf{x} = (T_{\mathrm{eff}}, [\mathrm{Fe}/\mathrm{H}], \mathrm{K_s}, \varpi)$, which are, respectively, the effective temperature, metallicity, apparent magnitude in the 2MASS $\mathrm{K_s}$~band, and the parallax.
These are chosen to mimic what one can do when combining a spectroscopic survey with the 2MASS photometry and \textit{Gaia} parallaxes.
Since these observables are insufficient for accurate age estimates of giant stars, the main tests are carried out using only stars in the general turnoff area of the HR diagram.
Giant stars are only considered in section \ref{sec:seismic_info} where we test the addition of seismic observables in the fitting step.

\subsection{Synthetic sample creation} \label{sec:synth_create}
All synthetic samples are created following the same procedure.
First, the desired age-metallicity distribution is defined on the discrete values available in the isochrone grid.\footnote{The grid resolution varies slightly between tests with ages in steps of $0.2$ to $0.4$~Gyr and metallicities in steps of $0.05$ to $0.1$~dex.}
Then the desired number of age-metallicity pairs (isochrones) is drawn following this distribution, and each one is assigned a mass in the valid interval for the chosen isochrone following a Salpeter initial mass function (IMF).
Low-mass dwarfs (below roughly 0.8$M_{\odot}$) are excluded by setting a metallicity-dependent lower limit on the main-sequence temperature given by $T_{\mathrm{eff,min}} = 5300-500\times[\mathrm{Fe}/\mathrm{H}]$.
Additionally, extra giant stars are added to the sample by drawing more models with a lower mass limit corresponding to the base of the red giant branch.
This is done to more closely mimic observational samples where giants are over-represented, compared to a sample following the IMF, owing to their high luminosity.

Having thus defined the true intrinsic parameters for each star, their parallaxes and apparent magnitudes are defined.
The parallaxes are either all set to the same value (cluster-like) or drawn from a continuous distribution (field-like).
The field-like distribution is based on the parallaxes in the magnitude-limited sample of \textit{Gaia} DR2 stars \citep{2018A&A...616A...1G} with metallicities based on SkyMapper photometry \citep{2019MNRAS.482.2770C}.
This sample is cut down to the roughly 6 million stars with a relative parallax uncertainty less than 10 per cent, and the stars are binned by their absolute J magnitudes, with bin width 0.5 magnitude.
In each bin the parallax distribution is approximately lognormal.
This is used to assign a parallax to each of the synthetic stars by drawing it from the lognormal distribution corresponding to its absolute J magnitude.
The apparent magnitudes are then calculated based on the parallax.

Finally, synthetic observations are drawn by perturbing the true parameters by an assumed uncertainty.
This is achieved by drawing the observed parameter from a normal distribution with the true parameter as the mean and the uncertainty as the standard deviation.
In the simplest case, the uncertainties are defined as a constant value, or a constant fraction of the parameter value, for all parameters.
However, in reality the parallax and magnitude uncertainties depend on the apparent magnitude.
This is taken into account in a similar way as for the parallax distribution, i.e. lognormal uncertainty distributions for the parallax and magnitudes are defined based on the SkyMapper sample in a number of bins in apparent magnitude.
These distributions are then used to draw a parallax and magnitude uncertainty, for each synthetic observation, based on the apparent magnitude.

\begin{table}
\caption{Uncertainties assigned to the synthetic stellar parameters in four different levels (0--3).
The final column indicates whether the true stellar parameters, drawn from the isochrones, are perturbed according to the parameter uncertainties before estimating individual ages.
See section~\ref{sec:synth_create} for an explanation of the field-like distributions and uncertainties.}
\begin{tabular}{llllllll}
\hline
Level & [Fe/H] (dex) & $T_{\mathrm{eff}}$ (K) & $\mathrm{K_s}$ & $\varpi$   &  Perturbed \\ \hline
0     & 0.05         & 50                     & 1\%            & 2\%        & No               \\
1     & 0.05         & 50                     & 1\%            & 2\%        & Yes              \\
2     & 0.10         & 100                    & Field-like     & Field-like & Yes              \\
3     & 0.15         & 150                    & Field-like     & Field-like & Yes              \\ \hline
\end{tabular}
\label{tab:unc}
\end{table}

\begin{figure}
\centering
\includegraphics[width=\columnwidth]{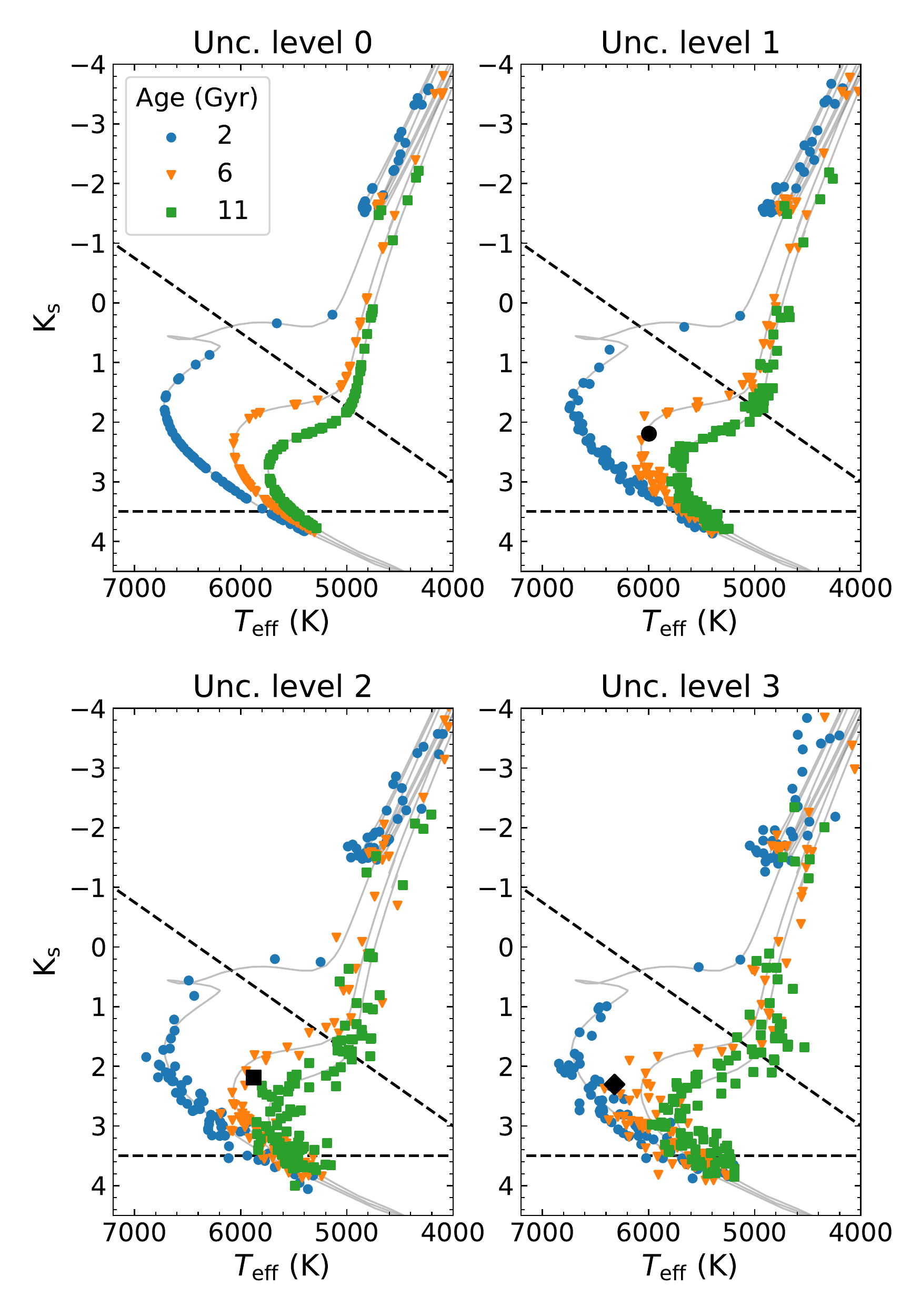}
\caption{Synthetic samples with $[\mathrm{Fe}/\mathrm{H}] = 0$~dex and ages of 2, 6, and 11~Gyr.
Each panel shows samples drawn with a different uncertainty level (see Table~\ref{tab:unc}).
The solid grey lines show the isochrones from which the samples are drawn, and the black dashed lines indicate the region used when selecting turnoff stars.
The black markers correspond to the same star at three different uncertainty levels for which $\mathcal{G}$~functions are shown in \autoref{fig:gfuncs_ex}.}
\label{fig:synth_HR}
\end{figure}

\begin{figure*}
\centering
\includegraphics[width=\textwidth]{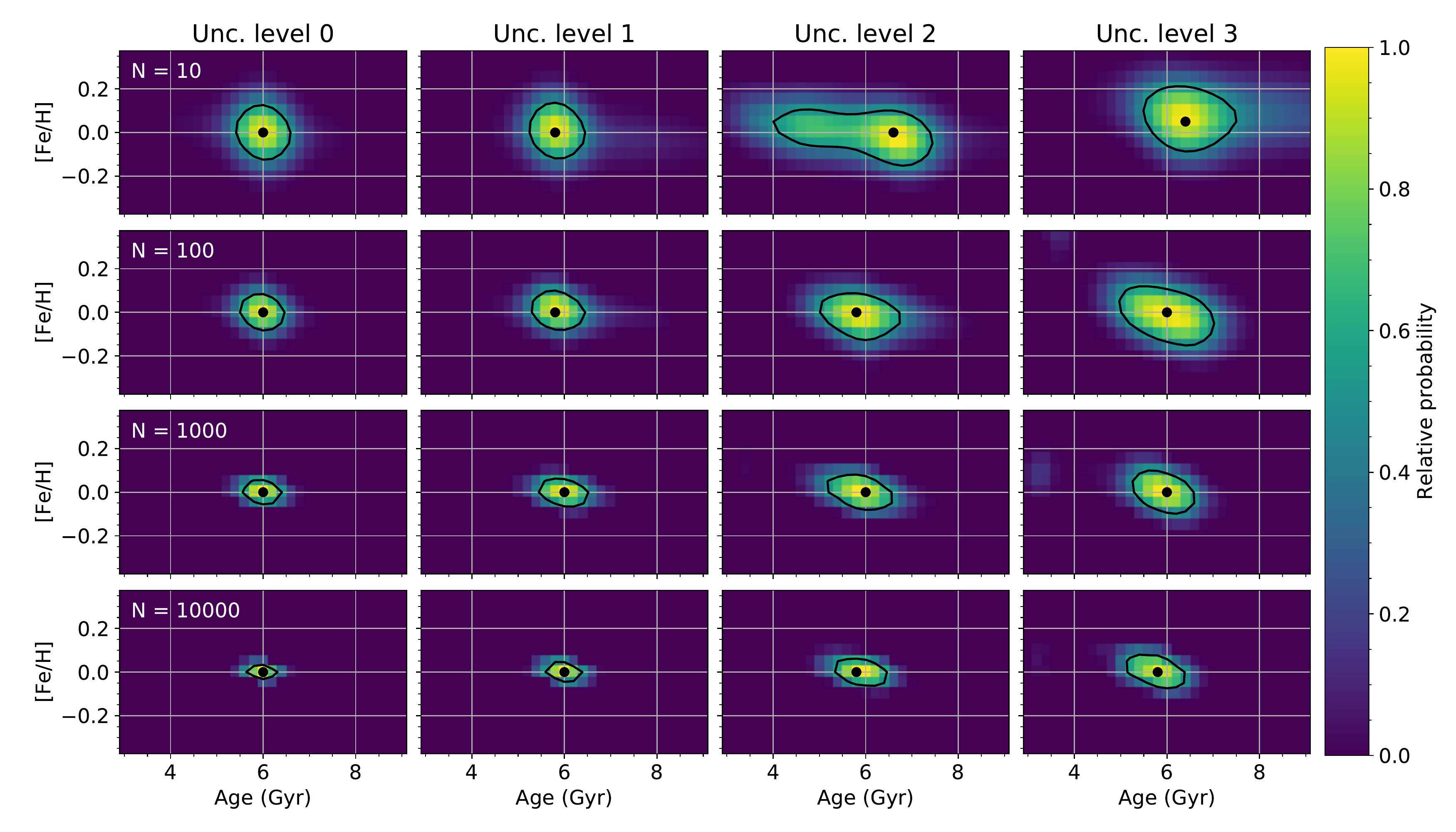}
\caption{SAMDs for the synthetic populations drawn from an isochrone with $[\mathrm{Fe}/\mathrm{H}] = 0$~dex and an age of 6~Gyr.
The SAMDs are all normalized to a maximum value of 1.
The columns show results for different uncertainty levels, and the rows for different sample sizes.
In each panel, the black dot marks the maximum of the SAMD and the black line marks the 0.5 contour level.}
\label{fig:single_iso_samds}
\end{figure*}

\begin{figure*}
\centering
\includegraphics[width=\textwidth]{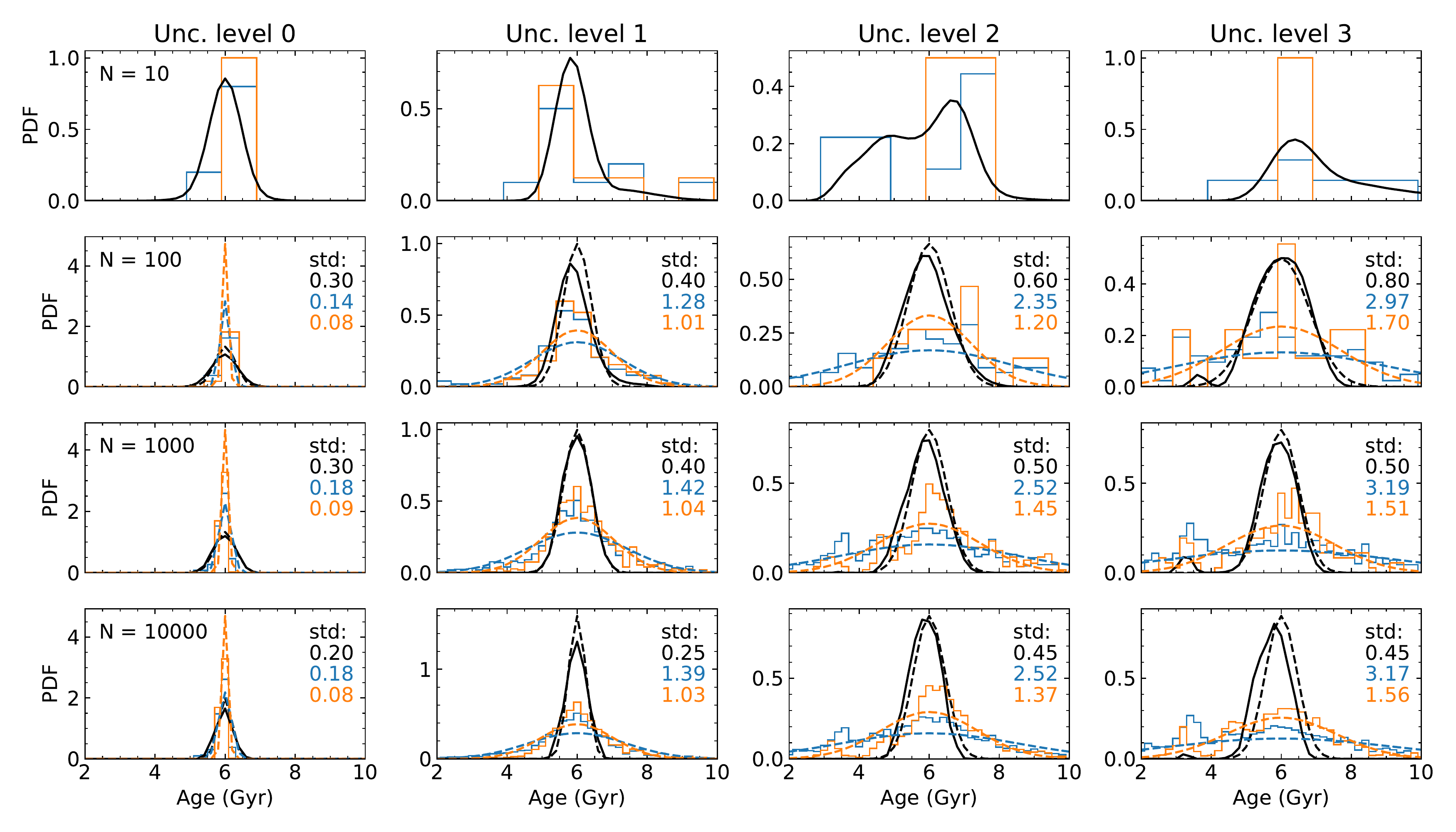}
\caption{SADs (black solid lines) and distributions of individually estimated ages (coloured histograms) for the same populations as shown in \autoref{fig:single_iso_samds}.
The blue histograms are made from the entire sample and the orange histograms are made from stars with a relative age uncertainty below 25 per cent.
For each distribution the standard deviation is printed in the panel, and a Gaussian with this standard deviation and a mean of 6~Gyr is shown as a dashed line of the corresponding colour.}
\label{fig:single_iso_sads}
\end{figure*}

\begin{figure}
\centering
\includegraphics[width=\columnwidth]{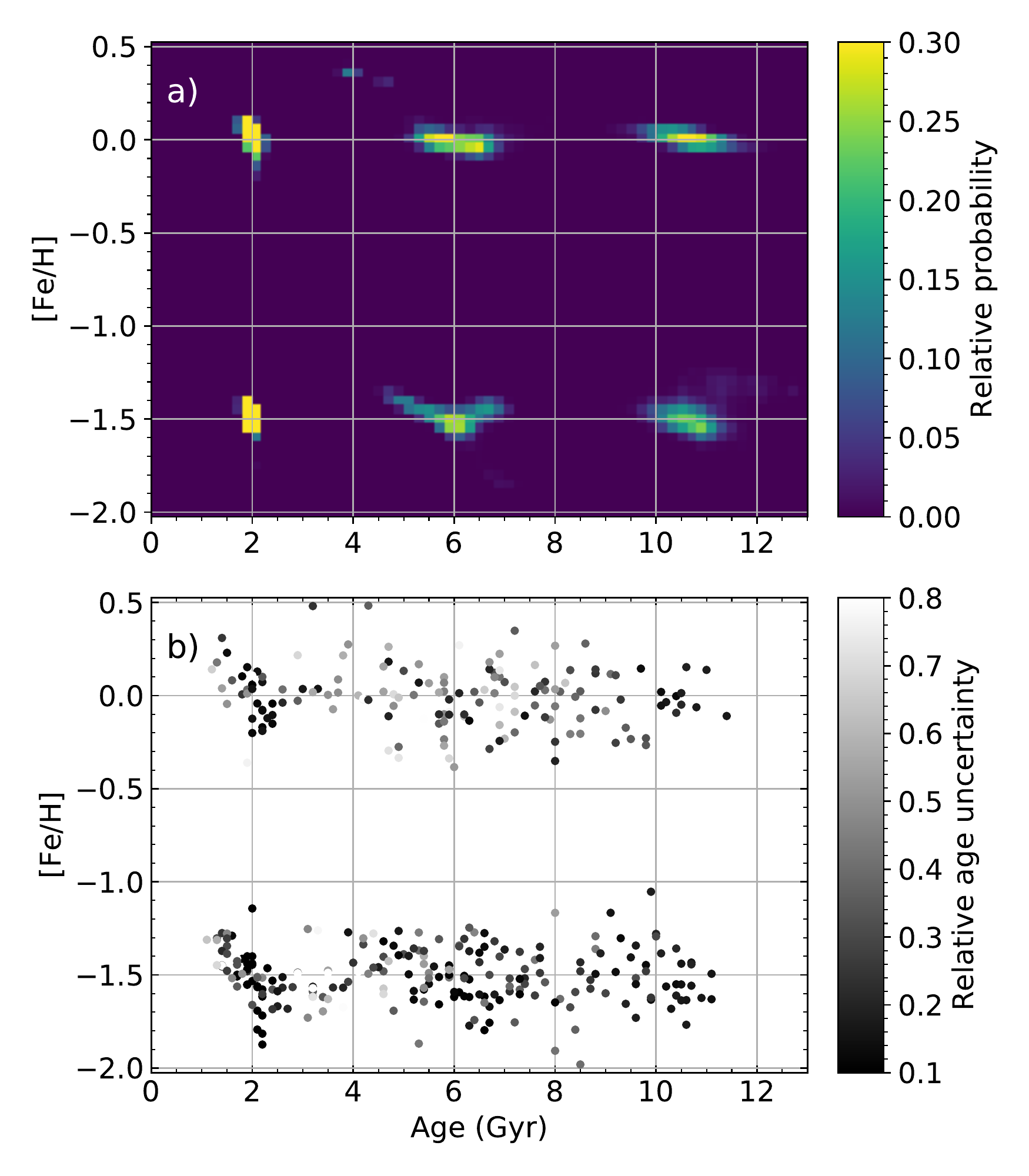}
\caption{Age-metallicity distributions for a synthetic sample made up of 100 stars drawn from each of six isochrones (600 stars in total).
Three of the isochrones have $[\mathrm{Fe}/\mathrm{H}] = 0$ and the other three have $[\mathrm{Fe}/\mathrm{H}] = -1.5$.
For each metallicity, the three isochrones have ages of 2, 6, and 11~Gyr.
a) SAMD normalized to a maximum value of 1.
The color scale is maximized at 0.3 to make the older populations more visible.
b) Metallicities versus individually estimated ages with the color indicating the relative age uncertainty.}
\label{fig:single_iso_samds_comb}
\end{figure}

\subsection{Mono-age populations} \label{sec:mono-age_pops}
The first and simplest test of the method is to apply it to a sample drawn from a single isochrone so all stars have the same true age and metallicity.
The aim is first of all to confirm that the method is able to recover the correct age and metallicity under ideal conditions, and then to test the accuracy of the recovered age distribution with different sample sizes and levels of uncertainties on the stellar parameters.

In the following tests, four different levels (0--3) of values and parameter uncertainties are used.
The parameter uncertainties are summarised in \autoref{tab:unc}.
For levels 0 and 1, all stars are placed at a distance of 1~kpc and assigned the same parameter uncertainties.
The difference is that for level 0 the true parameters are not perturbed, i.e. these samples follow the isochrone exactly with no scatter.
Each parameter still has an associated uncertainty since this is necessary to obtain a meaningful isochrone fit.
The purpose of this level is to test that we recover the input age and metallicity when the exact stellar parameters are known.
For levels 2 and 3, the field-like parallax distribution is used together with the corresponding parallax and magnitude uncertainties.
These two levels represent the most realistic uncertainties for a sample of field stars in the Galaxy.

For each uncertainty level, six samples have been defined: three with ages of 2, 6, and 11~Gyr at $[\mathrm{Fe}/\mathrm{H}] = 0$ and three with the same ages but $[\mathrm{Fe}/\mathrm{H}] = -1.5$.
For each of these combinations, a synthetic sample of 25000 stars has been drawn to make sure that each sample has at least 10000 turnoff stars.
\autoref{fig:synth_HR} shows example HR diagrams for random subsamples of 100 stars from each of the three samples with $[\mathrm{Fe}/\mathrm{H}] = 0$.

Starting with just the samples with $[\mathrm{Fe}/\mathrm{H}] = 0$ and $\tau = 6$ Gyr, the 2D $\mathcal{G}$~functions are calculated for the turnoff stars and the SAMD is estimated for different sample sizes.
Turnoff stars are selected from within the region between the dashed lines in the HR diagrams of \autoref{fig:synth_HR}.
This region was determined by eye to exclude stars on the lower main sequence and the giant branch where age uncertainties are largest.
In this test $\alpha = 250$ is used in the calculation of every SAMD so that any differences in the results are due only to the change of uncertainty level or sample size.
In general, $\alpha$ should be set to different values for different samples as explained in Appendix~\ref{sec:appB}.

The SAMDs are shown in \autoref{fig:single_iso_samds} for every combination of uncertainty level and sample size.
Ideally, the SAMD should be 1 in the bin centered on the true age and metallicity of the sample (6~Gyr and 0~dex) and 0 everywhere else.
In practice this is not possible, but for a sample of $10000$ stars with uncertainty level 0 (no parameter perturbation), the age and metallicity is recovered accurately as a narrow peak in the SAMD centered on the true values.
The distribution remains accurate as the sample size is decreased all the way down to 10 stars.
At the same time, the precision decreases making the peak wider.
The same trends are seen for uncertainty levels 1, 2, and 3 for all samples with 100 stars or more.
The peak of the SAMD is within 0.2 Gyr (the grid resolution) of the true age in all these cases and the precision decreases with increasing uncertainty and decreasing sample size.
For the highest uncertainties, the correlation between metallicity and age is also visible in the SAMD contours.
Only when reaching a sample size of 10 does the peak of the SAMD fail to recover the true age and metallicity accurately for uncertainty levels 2 and 3.

Individual ages for the stars can be estimated by locating the mode of each individual $\mathcal{G}$~function as shown in \autoref{fig:gfuncs_ex}.
Doing this for each star in every sample, the distribution of individually estimated ages can be compared with the SAD (the SAMD summed over the metallicity at each age) as shown in \autoref{fig:single_iso_sads}.
In general, the SAD is found to recover the true age distribution (a delta function in this case) more precisely than the distribution of individual ages.
However, for uncertainty level 0 the distributions of individual ages are more precise than the SADs.
This is likely because almost all individual $\mathcal{G}$~functions peak at the true age of 6~Gyr even though they have a finite width that propagates into the SAMD.
For the higher uncertainty levels, where the true parameters are actually perturbed before calculating the $\mathcal{G}$~function, the SAD has a lower standard deviation than the distribution of individually estimated ages.
The standard deviation of the SAD also decreases as the sample size increases which is not the case for the individual age distributions.
Of course, the standard deviation of the SAD depends on the amount of smoothing, i.e. the value of $\alpha$, but we have no reason to believe, from looking at \autoref{fig:single_iso_samds}, that insufficient smoothing has been applied.
Changing the value of $\alpha$ also does not change the fact that the SAD becomes more precise with increasing sample size.

In addition to the decreased precision when the uncertainty level is increased, the SAD also becomes slightly less accurate.
There is a tendency for the peak of the SAD to underestimate the true age which is seen most clearly at uncertainty level 3 with $N=10000$.
The cause of this underestimation is not obvious, but it is likely related to the complicated transformation between stellar surface parameters and age.
We see in \autoref{fig:synth_HR} that the star highlighted as an example (with black symbols) scatters to a lower and higher effective temperature for uncertainty levels 2 and 3, respectively.
As shown in \autoref{fig:gfuncs_ex} this leads the age to be overestimated in the first case and underestimated in the second.
However, because the isochrones of different ages are more separated in the HR~diagram at higher temperatures, the $\mathcal{G}$~function is narrower in the case where the temperature is overestimated.
In this way, it is likely that the single stars with underestimated ages, and a more localised $\mathcal{G}$~function, have a greater impact on the shape of the SAMD.
In the same vein, some stars will have a bimodal $\mathcal{G}$~function due to isochrones of different age-metallicity combinations overlapping.
This is likely the cause of the secondary peak between 2 and 4~Gyr seen in the distributions of individually estimated ages at the highest uncertainty levels.

Next, we fix the sample size to 100, use uncertainty level 3, and consider all six age and metallicity combinations defined above.
\autoref{fig:single_iso_samds_comb}a shows the estimated SAMD when giving all six samples as input at the same time.
All of the samples are easily identified in the SAMD, and they all peak within 0.5~Gyr of their true age.
The oldest populations are recovered with the lowest precision because the individual $\mathcal{G}$~functions are less precise.
Even though this makes the younger peaks higher, the total fraction of the SAMD contained in each peak are all very similar.
This fraction, obtained by integrating the SAMD in regions around the peaks, spans values from $14.5\%$ (for $[\mathrm{Fe}/\mathrm{H}] = -1.5$, 11~Gyr) to $18.4\%$ (for $[\mathrm{Fe}/\mathrm{H}] = -1.5$, 6~Gyr).
So the fraction of the SAMD in each population is within about $15\%$ of the fractional number of stars ($1/6\approx16.7\%$) in that population.

The SAMD can be compared with the distribution of the individually estimated ages and metallicities (\autoref{fig:single_iso_samds_comb}b).
Only the youngest populations are easily identified among the individual estimates.
The older populations only stand out clearly if all stars with age uncertainties greater than 25 per cent are removed -- this is 70 per cent of the sample.

\begin{figure*}
\centering
\includegraphics[width=\textwidth]{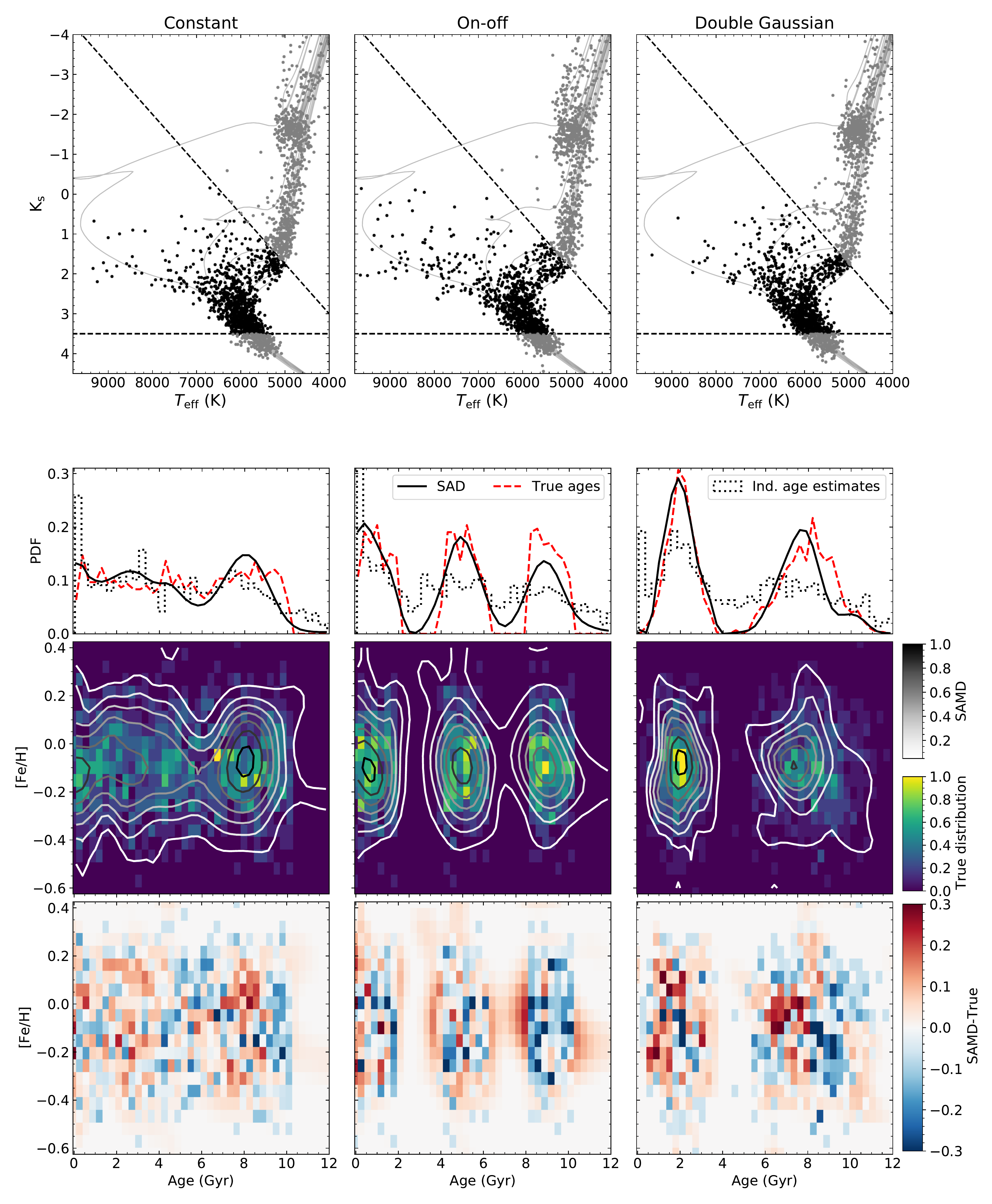}
\caption{HR diagrams, SAMDs, and SADs for three synthetic samples with different age distributions.
Each sample consists of 1000 turnoff stars highlighted in black in the HR diagrams in the upper panels.
In the middle panels, the true age-metallicity distributions are plotted as 2D-histograms and the estimated SAMDs are overplotted as greyscale contours.
The histograms of true values use the same binning as the SAMDs.
The panels just above show the SADs (black lines), true age distributions (red dashed lines) and histograms of individually estimated ages (black dotted lines).
The difference between the SAMD and the true distribution, after normalizing each distribution to have an integral equal to 1, is shown in the lower panels.}
\label{fig:samd_and_ages_to}
\end{figure*}

\subsection{Populations with extended age distributions} \label{sec:extended_age_pops}
We now test the method on samples with extended age distributions.
In all following tests, the samples are defined with uncertainty level 2.
The true metallicities are drawn from a normal distribution, independent of the age, with a mean of $-0.1$~dex and a standard deviation of 0.15~dex.
Three different age distributions are tested: 1) a flat distribution of ages below 10~Gyr, 2) an on-off distribution with ages distributed equally between and within the intervals 0 to 2, 4 to 6, and 8 to 10~Gyr, 3) a distribution given by the sum of two Gaussians
\begin{equation}
    p_{\mathrm{gauss}}(\tau) = 1.8f(\tau;\mu=2,\sigma=0.6) + f(\tau;\mu=8,\sigma=1.2) \; ,
\end{equation}
with $\tau$ in Gyr.
The first two distributions were used by \citet{2005ESASP.576..171J} because they look very similar in an HR diagram despite their significant differences.
The last one is inspired by \citet{1999MNRAS.304..705H} who tested their method on an almost identical distribution.

\autoref{fig:samd_and_ages_to} shows the three samples in HR diagrams as well as the SAMD obtained using 1000 turnoff stars.
The turnoff stars are selected from the larger samples shown in the HR diagrams (including giants) which are drawn from the age distributions given above.
Therefore, the turnoff stars do not follow these age distributions exactly.
Qualitatively, the estimated SAMDs recover the true underlying distributions very well.
One is easily able to distinguish the constant and on-off age distributions in the age-metallicity space, and the on-off distribution shows three peaks on top of the corresponding peaks in the true distribution.

The difference between the SAMD and the true distribution in the bottom panels of \autoref{fig:samd_and_ages_to} reveal a few areas where the SAMD is either over- or underestimated systematically across several adjacent age-metallicity bins.
In the oldest of the two Gaussian peaks, for example, the SAMD has its local maximum at around 7.5~Gyr instead of the true value of 8~Gyr.
As a result, the SAMD is overestimated in the region around this maximum and underestimated in the older regions.
We believe this is related to the tendency of the SAMD to slightly underestimate ages as observed also for the mono-age populations in \autoref{fig:single_iso_samds} and discussed in \autoref{sec:mono-age_pops}.

When marginalizing the SAMDs over metallicity to get the SAD, the true age distribution is seen to be well reproduced.
In all of these cases, the SAD is a better approximation of the true age distribution than the histogram of individual ages which have been estimated using the modes of the individual PDFs.
Especially the on-off distribution is poorly recovered by the individual age distribution which looks more or less constant for ages from 2 to 10~Gyr.
In total, 61 per cent of the individual estimates are within the three regions where the true distribution is `on'.
For the SAMD, 83 per cent of the integrated probability is within these regions.
Of course, this comparison is not quite fair because in real applications one would make some quality cuts on the individual ages.
We take this into account and consider the comparison between age distributions more carefully in section~\ref{sec:compare_ages}.

\begin{figure*}
\centering
\includegraphics[width=\textwidth]{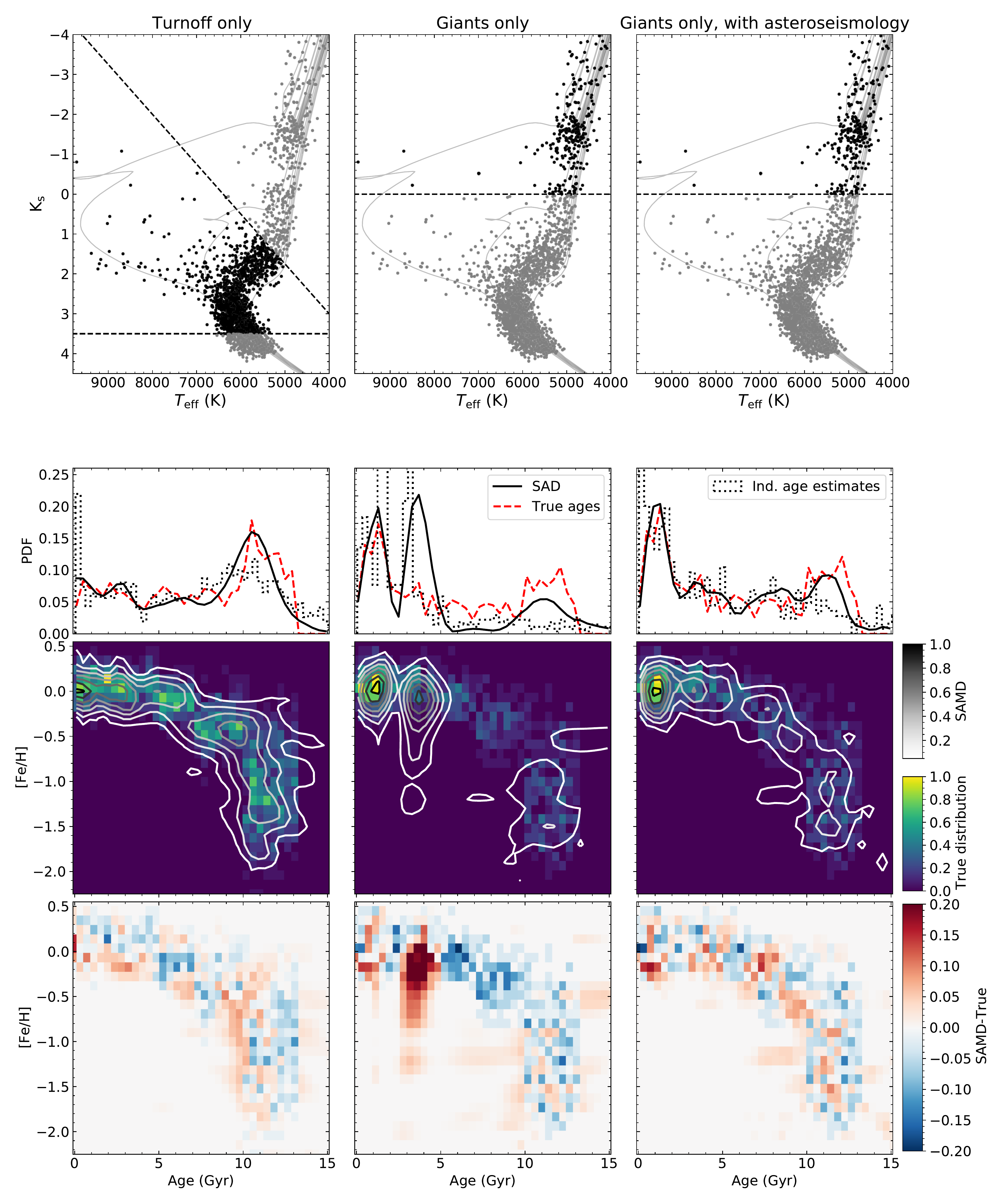}
\caption{Like \autoref{fig:samd_and_ages_to} but each column shows a different selection stars from the Milky Way-like sample.
From left to right: turnoff stars only, giants only, and giants only with asteroseismic parameters in the fit.}
\label{fig:samd_and_ages_MWmodel}
\end{figure*}

\begin{table}
\caption{Age ranges and metallicity distributions for the subpopulations making up the Milky Way-like population.}
\begin{tabular}{llr}
\hline
Population & Ages (Gyr) & {[}Fe/H{]} (dex) \\ \hline
1          & 0-5        & $0.00\pm0.15$    \\
2          & 5-7        & $-0.15\pm0.15$   \\
3          & 7-10       & $-0.40\pm0.20$   \\
4          & 10-13      & $-0.80\pm0.30$   \\
5          & 10-13      & $-1.40\pm0.30$   \\ \hline
\end{tabular}
\label{tab:MWpops}
\end{table}

\subsection{Milky Way-like population} \label{sec:MW_pop}
Finally, a sample with metallicity variations is tested.
We define what we refer to as a Milky Way-like population by using age-metallicity combinations inspired by the Besan\c{c}on Galaxy model \citep[][Table 1]{2003A&A...409..523R}.
The ages and metallicities are drawn from the subpopulations given in \autoref{tab:MWpops} to get 100 turnoff stars for each Gyr for each population (meaning 500 from population 1, 200 from population 2, etc. giving 1600 stars in total).
The ages are drawn from uniform distributions within each population while the metallicities are drawn from normal distributions.
This is not meant to mimic the actual Milky Way age-metallicity distribution, but rather to provide a sample with somewhat realistic age-metallicity correlations.
It lacks correlations between parameters such as position in the Galaxy and metallicity, and the number of stars in each age interval is not representative of the true Galactic distribution.

The SAMD has been calculated for the entire sample of turn-off stars and the result is shown in the left-hand column of \autoref{fig:samd_and_ages_MWmodel}.
The estimated SAMD seems to be a good representation of the true distribution over most of the grid.
The biggest differences are seen near the edges of the true distribution where the residuals are largest.
Especially at the old edge of the grid where the SAMD is underestimated at all metallicities.
This is also seen in the SAD which peaks near 11 Gyr like the true distribution but fails to reproduce the sharp transitions at 10 and 13~Gyr.
However, it is a clear improvement over the distribution of individual age estimates which shows a broader peak with a tail going all the way up to the edge of the grid (20~Gyr in this case, outside of the figure).
The agreement is better if these stars are excluded, but still not as good as the SAD (see section~\ref{sec:compare_ages}).

\subsection{Adding asteroseismic information} \label{sec:seismic_info}
So far we have considered turnoff stars only since the ages of giants stars are poorly constrained by isochrone fitting to their surface parameters.
For many giants (with ages above about 2~Gyr) the age estimate from isochrone fitting is underestimated.
This is seen clearly in the middle panel of \autoref{fig:samd_and_ages_MWmodel} where the SAMD is estimated from giant stars alone.
Only the youngest part of the distribution (below about 2.5~Gyr) is recovered in the SAMD.
A secondary peak, which does not exist in the true distribution, is found for ages between 2.5 and 5~Gyr.
This peak is caused by the ages of most older stars being underestimated.

It is possible to put much stronger constraints on the ages of giants if asteroseismic data are available.
To test the extent to which seismic observables improve the SAMD estimate for giants, we add the seismic parameters $\Delta\nu$ and $\nu_{\mathrm{max}}$ to the stellar models using the scaling relations \citep{1986ApJ...306L..37U, 1995A&A...293...87K}
\begin{align}
  \frac{\Delta\nu}{\Delta\nu_{\sun}} &=
  \left(\frac{M}{M_{\sun}}\right)^{1/2}
  \left(\frac{R}{R_{\sun}}\right)^{-3/2} \label{eq:dnu_scal} \; ,
  \\
  \frac{\nu_{\mathrm{max}}}{\nu_{\mathrm{max},\sun}} &=
                          \left(\frac{M}{M_{\sun}}\right)
                          \left(\frac{R}{R_{\sun}}\right)^{-2}
                          \left(\frac{T_{\mathrm{eff}}}
                          {T_{\mathrm{eff},\sun}}\right)^{-1/2}
                          \label{eq:numax_scal} \; .
\end{align}
These parameters are then drawn along with the surface parameters when the synthetic sample is created.
The relative uncertainties assigned to the seismic parameters are 2 per cent for $\Delta\nu$ and 4 per cent for $\nu_{\mathrm{max}}$.
These are larger than the typical uncertainties for Kepler red giants \citep{2018ApJS..236...42Y} but smaller than the mean uncertainties for the first TESS seismic data \citep{2020ApJ...889L..34A}.

The right-hand panel of \autoref{fig:samd_and_ages_MWmodel} is based on the same sample as the middle one, but now with the seismic parameters included in the fitting step.
Now the age-metallicity distribution is recovered just as well as it was for the turnoff sample (although the distribution is different since we use a different selection of stars).
Looking at the age distribution, the biggest difference from the true distribution is near the old edge where the SAD only shows a narrow bump and falls off too early.
But, again, the SAD is an improvement over the distribution of individually estimated ages which does not show a strong indication of a bump between 10 and 13~Gyr.

\begin{figure}
\centering
\includegraphics[width=\columnwidth]{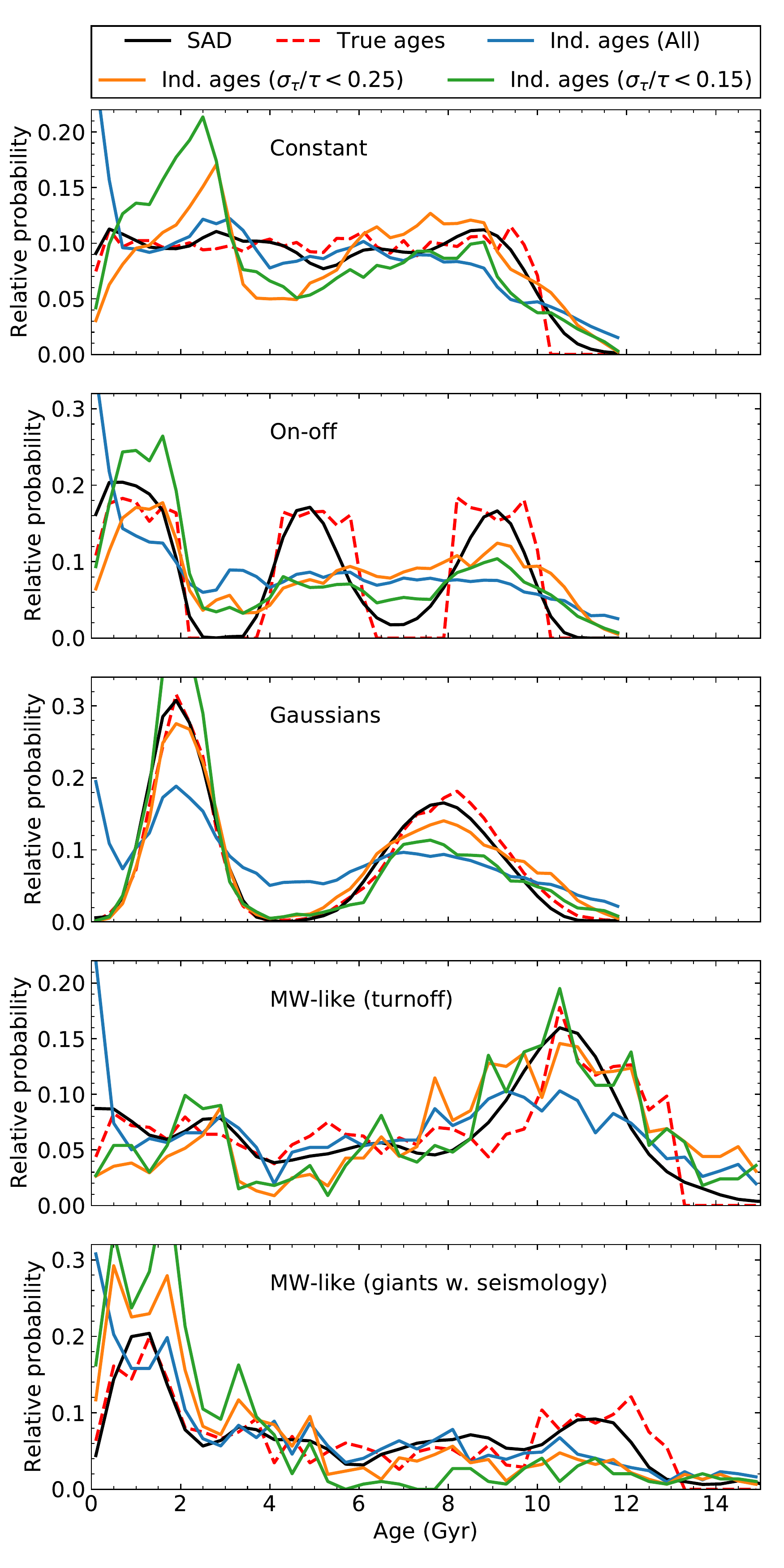}
\caption{True and recovered age distributions for five of the samples shown in Figures \ref{fig:samd_and_ages_to} and \ref{fig:samd_and_ages_MWmodel} (excluding giants without seismology).
For the three non-MW-like samples, we here show the results for sample sizes of 10000 turnoff stars (which can be compared with \autoref{fig:samd_and_ages_to} where the samples contain 1000 turnoff stars).
Recovered age distributions are shown in the form of SADs and individual age distributions, and the latter is shown with all stars and two different cuts on the relative age uncertainty.}
\label{fig:ages_1D_comparison}
\end{figure}

\begin{figure}
\centering
\includegraphics[width=\columnwidth]{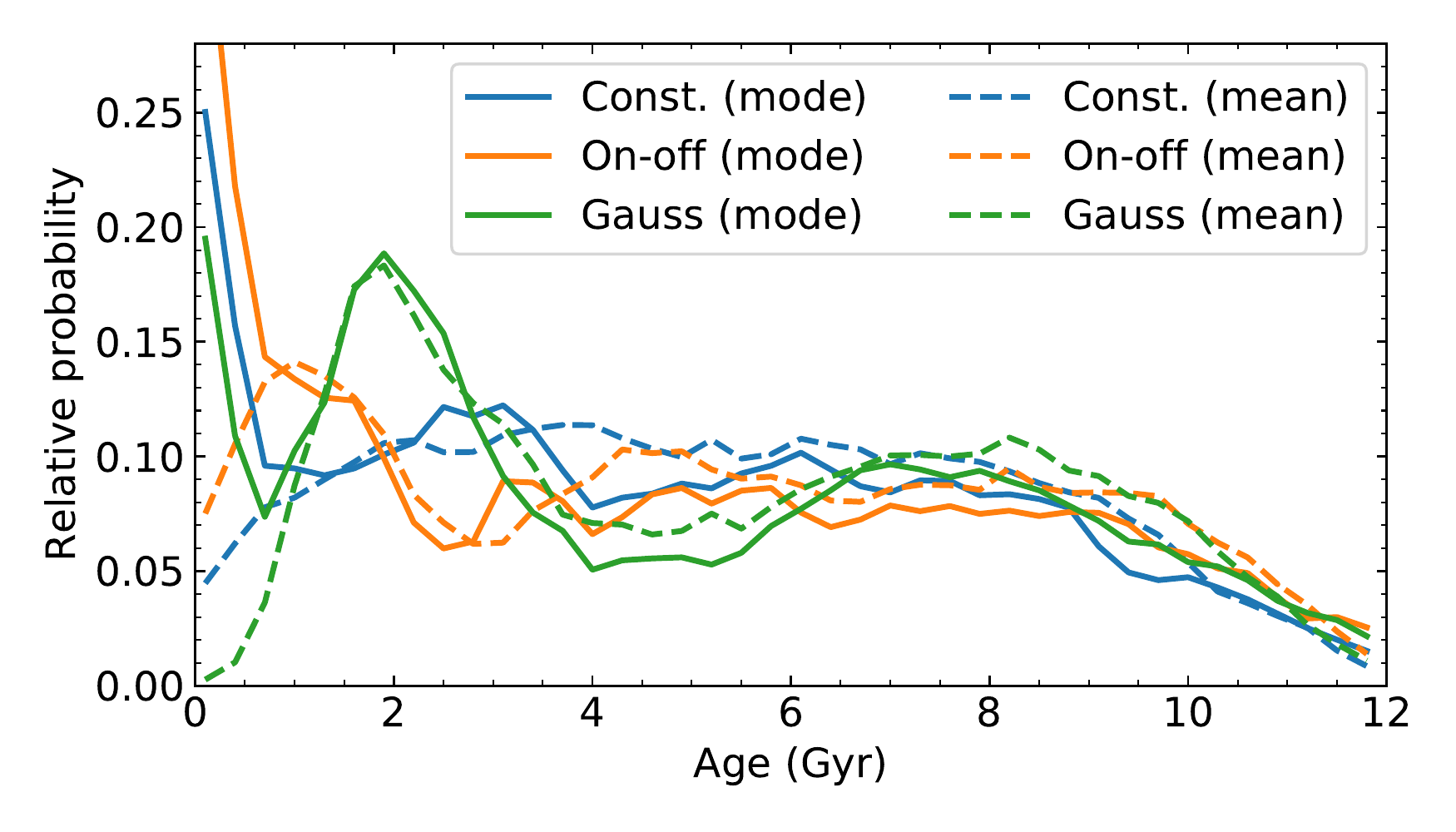}
\caption{Distributions of individual age estimates (including all stars) for the constant, on-off, and double Gaussian samples with $N = 10000$.
Estimates using the modes of the $\mathcal{G}$~functions are shown as solid lines and those using the mean as dashed lines.}
\label{fig:ages_1D_mode_mean}
\end{figure}

\begin{table*}
\caption{Root mean square of the differences between the true and recovered age distributions shown in \autoref{fig:ages_1D_comparison} (and in \autoref{fig:samd_and_ages_to} for the samples with $N = 1000$).}
\begin{tabular}{llllllll}
                            & \multicolumn{7}{c}{RMS of difference from the true age distribution}          \\ \cline{2-8} 
                            & \multicolumn{1}{c}{} & \multicolumn{2}{c}{Ind. ages (All)} & \multicolumn{2}{c}{Ind. ages ($\sigma_{\tau}/\tau < 0.25$)} & \multicolumn{2}{c}{Ind. ages ($\sigma_{\tau}/\tau < 0.15$)} \\
Sample                      & SAD                  & Mode             & Mean             & Mode                   & Mean                   & Mode                   & Mean                   \\ \hline
Constant ($N = 1000$)         & 0.17                 & 0.27             & 0.17             & 0.23                   & 0.18                   & 0.28                   & 0.20                   \\
Constant ($N = 10000$)        & 0.12                 & 0.26             & 0.18             & 0.23                   & 0.18                   & 0.28                   & 0.23                   \\
On-off ($N = 1000$)           & 0.25                 & 0.48             & 0.40             & 0.42                   & 0.40                   & 0.44                   & 0.40                   \\
On-off ($N = 10000$)          & 0.22                 & 0.50             & 0.43             & 0.41                   & 0.38                   & 0.42                   & 0.39                   \\
Gaussians ($N = 1000$)        & 0.20                 & 0.41             & 0.36             & 0.18                   & 0.22                   & 0.30                   & 0.25                   \\
Gaussians ($N = 10000$)       & 0.13                 & 0.38             & 0.33             & 0.15                   & 0.19                   & 0.28                   & 0.24                   \\
MW-like (turnoff, $N = 1600$) & 0.15                 & 0.25             & 0.18             & 0.22                   & 0.22                   & 0.20                   & 0.19                   \\
MW-like (giants, $N = 1600$)  & 0.14                 & 0.32             & 0.22             & 0.31                   & 0.21                   & 0.46                   & 0.36                   \\ \hline
\end{tabular}
\label{tab:ages_RMS}
\end{table*}

\subsection{Comparing age distributions} \label{sec:compare_ages}
Having shown that the SAMD can recover different age-metallicity distributions fairly closely, it is interesting to take a closer look at the age distributions.
Since we are working with synthetic samples, avoiding all systematic errors relating to the models, the conditions are optimal for the SAD to reproduce the true age distribution.
But this is also true for the distributions of individually estimated ages; therefore, it is interesting to test whether the SAD is a better approximation to the true age distribution than the distribution of individual ages.
If this is the case for synthetic samples, it should also hold true when applied to real data.

When estimating the SAMD, we make use of the entire $\mathcal{G}$~function for each star.
This is one of the strenghts of the method.
When estimating individual ages, one needs to choose a statistic like the mode or the mean of the $\mathcal{G}$~function.
Then associated uncertainties can be calculated in order to select subsamples with only the most precise age estimates.

In \autoref{fig:ages_1D_comparison} the true age distributions for the five samples with well-recovered SAMDs are shown along with the SAD and distributions of individual ages.
For the two Milky Way-like samples, the data are the same as in \autoref{fig:samd_and_ages_MWmodel}.
For the three other samples, \autoref{fig:ages_1D_comparison} shows the results based on sample sizes of 10000 stars.
Compared with the SADs in \autoref{fig:samd_and_ages_to}, which are based on sample sizes of 1000 stars, it is clear that the increase in sample size also increases the accuracy of the SAD.

When including all individual ages, the distributions all show a peak at $\tau = 0$.
This is caused by those stars for which the mode of the $\mathcal{G}$~function is at the edge of the age grid.
By using the mean of the $\mathcal{G}$~function as the age estimate this peak disappears (see \autoref{fig:ages_1D_mode_mean}), but then the ages of some of the stars which really are that young are overestimated, leading instead to a small dip towards the edge of the grid.
For higher ages, where the grid edge effects play no role, the mode and mean ages are very similar.

When excluding stars with relative age uncertainties above 25 or 15 per cent (which also removes those stars with a $\mathcal{G}$~function peaking at the grid edge), the agreement with the true age distribution is not necessarily improved.
Only for the double Gaussian distribution do we see a clear improvement when limiting the sample to those stars with $\sigma_{\tau}/\tau < 0.25$.
But with the stronger restriction of $\sigma_{\tau}/\tau < 0.15$, stars are mostly removed from the older peak and the apparent relative number of stars in the two peaks becomes skewed.
The relative number of young stars is also increased beyond the true value for the constant age distribution; in this case, the best agreement is found without cutting out any stars at all.

The agreement of the different distributions with the true age distribution is quantified in \autoref{tab:ages_RMS} which lists the RMS differences.
For all but one sample (Gaussians, $N=1000$), the SAD has an equal or lower RMS difference than any of the distributions of individual ages.
These values also show that going from a cut in relative uncertainty of 25 to 15 per cent only makes the RMS difference larger since one then generally selects a larger fraction of younger stars.

\begin{figure*}
\centering
\includegraphics[width=\textwidth]{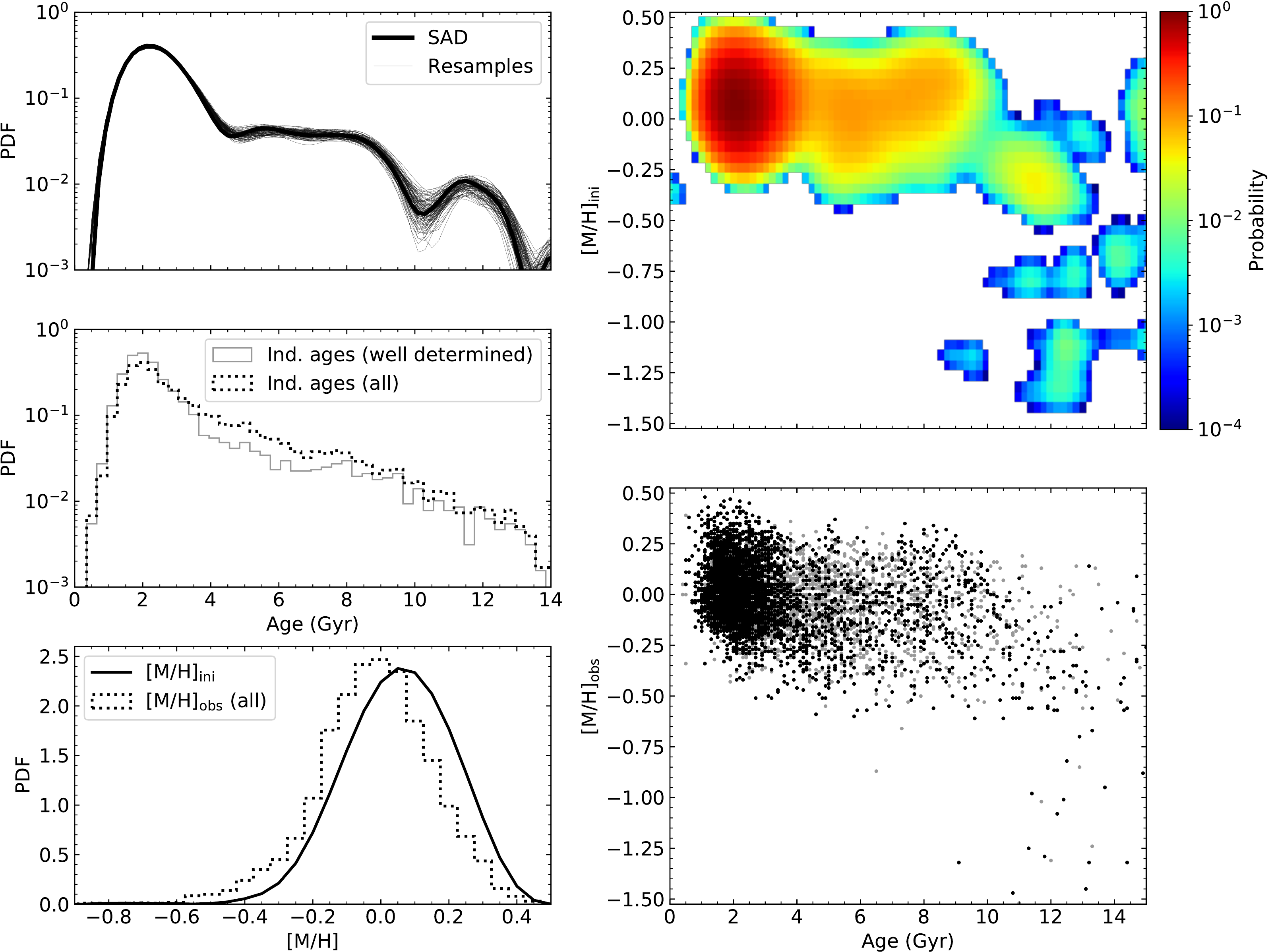}
\caption{Age-metallicity distribution for the 5976 stars in the Geneva-Copenhagen survey with temperatures estimated using the infrared flux method by \citet{2011A&A...530A.138C}.
On the right-hand side, the upper panel shows the SAMD and the lower panel shows the distribution with individual age estimates.
Individual age estimates meeting the requirements of being ``well determined'', as defined by \citet{2011A&A...530A.138C} ($\sigma_{\tau}/\tau < 0.25$ or $\sigma_{\tau}$ < 1~Gyr), are shown in black and the rest in grey.
On the left-hand side, the upper panel shows the SAD (SAMD marginalized over metallicity) and the middle panel shows the distributions of individual ages.
The SAD (in thick black) is shown along with 100 SADs obtained by running the analysis on random resamples of the data (in thin grey).
The lower panel shows the distributions of observed metallicities (used as input for the isochrone fitting) and the recovered distribution of initial metallicities given by marginalizing the SAMD over age.}
\label{fig:ages_gcs}
\end{figure*}

\begin{figure}
\centering
\includegraphics[width=\columnwidth]{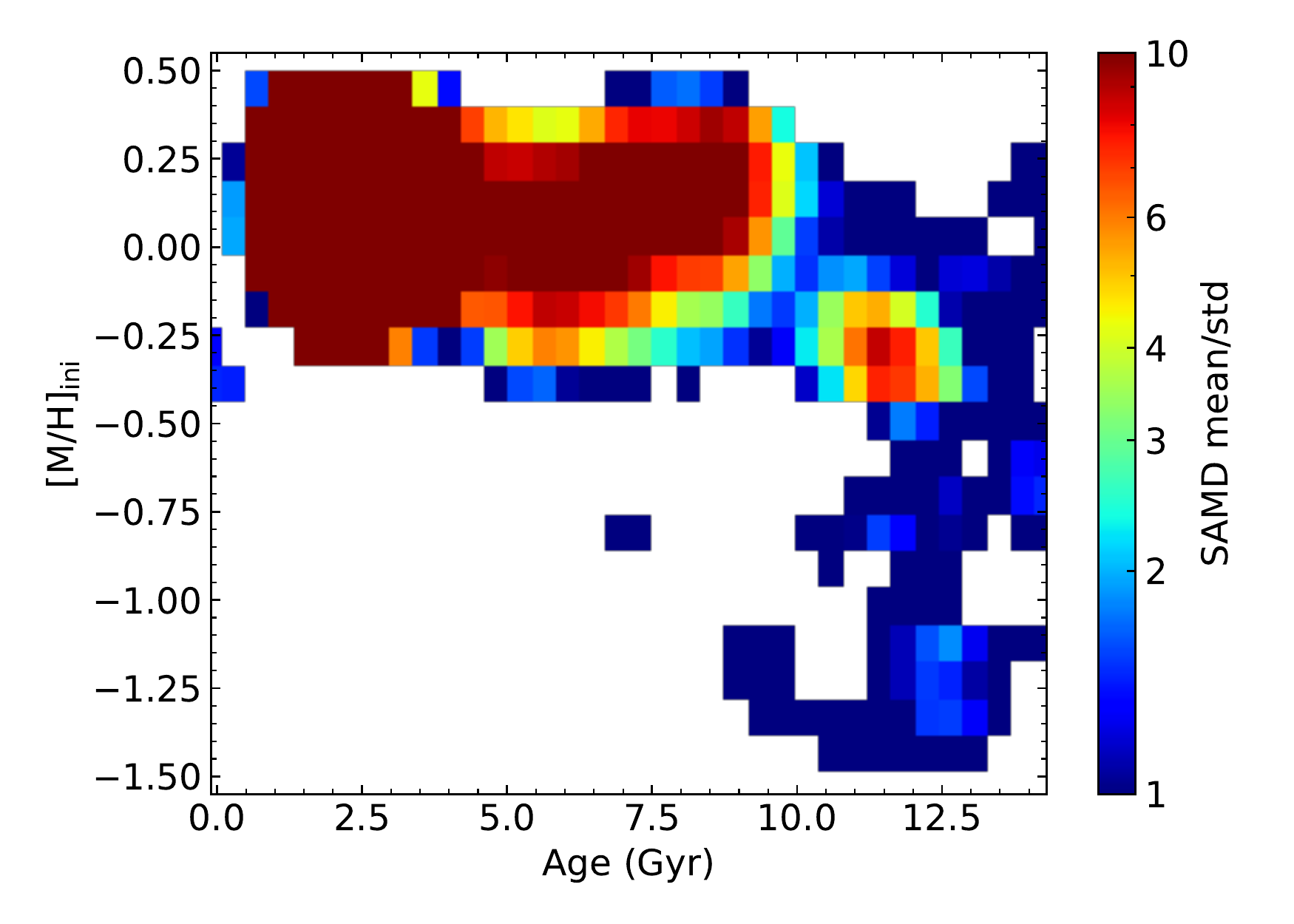}
\caption{An illustration of the signal-to-noise ratio (SNR) in the GCS SAMD.
It is calculated as the mean divided by the standard deviation (per age-metallicity bin) of 100 SAMDs obtained by running the analysis on random resamples of the data.
Peaks found in the SAMD in the blue regions of this plot (with SNR below 2) are deemed statistically insignificant.}
\label{fig:ages_gcs_unc}
\end{figure}

\section{Application to the Geneva-Copenhagen survey} \label{sec:real_data_test}
As a final demonstration of the method, we apply it to a sample of nearby dwarf stars from the Geneva-Copenhagen survey \citep[GCS;][]{2004A&A...418..989N}.
This is a magnitude limited sample of F and G dwarfs in the Solar neighbourhood (within about 200~pc) which was defined as a benchmark for studies of the Galactic disk.
Among other things, it has been used to study the age-metallicity relation of stars in the Solar neighbourhood.
Here this relation is re-derived in the form of the SAMD to see if any details are revealed beyond what is apparent when using individual estimates for the ages.

$T_{\mathrm{eff}}$ and [M/H] for the sample stars have been taken from the updated analysis by \citet{2011A&A...530A.138C}.
They restrict their age analysis to the subsample of 5976 stars for which $T_{\mathrm{eff}}$ was determined using the infrared flux method.
We make the same selection.
Then parallaxes and $\mathrm{K_s}$ magnitudes have been collected from the SIMBAD database, and each star has been fitted to the same isochrone grid as used for the synthetic samples.
Finally, the SAMD has been calculated using the entire sample and is shown in \autoref{fig:ages_gcs} along with an age-metallicity scatter plot of individual values.

Since we are now working with real stars, their initial metallicities are unknown.
Therefore, the fit is made comparing the observed metallicity with the actual surface metallicity of each individual model in the isochrone grid.
However, the result is expressed in terms of the initial model metallicity which is what defines each isochrone.
This causes a shift between the observed and reconstructed metallicity distribution with the latter being about 0.06~dex higher on average.
This difference simply reflects the decrease of the surface metallicity with time due to settling of heavy elements.

Overall, the age-metallicity relation found here looks very similar to the one found by \citet[][Fig. 16]{2011A&A...530A.138C}.
As they discuss, the sample consists mainly of young stars;
The SAMD indicates that about 75 per cent of the sample is younger than 4~Gyr.
This is caused by several factors, including the fact that older stars spend more time further away from the sampled volume (due to vertical oscillations in the disk) and that the magnitude limit of the survey leads to a bigger selection volume for younger, brighter stars.

At an age of about 10~Gyr, the SAMD shows a transition from a plateau of younger stars at solar metallicities to older stars at sub-solar metallicities.
This shows up as a minimum in the SAD which is not seen in the distribution of individual ages.
The significance of this minimum is debatable since the number of old stars is low.
Only 2.3 per cent of the sample is older than 10~Gyr, according to the SAMD, which corresponds to about 160 stars.
As shown earlier in the synthetic tests, a minimum of 100 stars per gigayear is desirable.
Additionally, the statistical uncertainty of the SAD is largest around an age of 10~Gyr as demonstrated by the scatter of the SADs obtained by running the analysis on 100 random resamples.
On the other hand, all of these resamples do show a local minimum even if it is shallow in some cases.
In any case, the SAMD indicates a transition at 10~Gyr which is not seen in the individual age estimates.

The SAMD shows a number of smaller peaks at metallicities below $-0.6$~dex.
Comparing with the plot of individually estimated ages, it is clear that they are mostly made up of one or a few stars.
To quantify their significance, we show a measure of the signal-to-noise ratio (SNR) of the SAMD, based on the resample analysis, in \autoref{fig:ages_gcs_unc}.
The SNR reaches values above 10 in the most densely populated regions of the age-metallicity space.
In the region below $-0.6$~dex, however, the SNR only reaches a maximum value of 2.
Thus, we conclude that the low-metallicity peaks in the SAMD are statistically insignificant.

An in depth analysis of the cause of the minimum at 10~Gyr goes beyond the scope of this paper; however, it is not unthinkable that it represent a real dip in the SFH.
It could be related to the recent finding that the last major merger with the Milky Way happened about 10~Gyr ago \citep{2018Natur.563...85H, 2019ApJ...881L..10S}.
That is not to say that the oldest GCS stars are accreted, but the merger event could have temporarily lowered the star formation rate in the Galactic disk.
The SFH could also be shaped by the gas infall history.
For example, the two-infall model predicts an SFH with an old peak related to the first infall episode separated from a continuous formation of younger stars in the second infall episode \citep{1997ApJ...477..765C, 2019A&A...623A..60S}.
Additionally, the selection function of the GCS sample as well as the fact that stars migrate in the disk (coming into the Solar neighbourhood from the inner and outer disk) surely play a role in the observed SFH.
The age-metallicity distribution of stars in the GCS sample has previously been reproduced with a chemical evolution model of the Galaxy including radial migration of stars and flow of gas \citep{2009MNRAS.396..203S}.
The same model was used to explain the alpha element dichotomy observed in the nearby stellar disc indicating that radial migration plays a significant role in shaping the solar neighbourhood.
\citet{2018ApJ...865...96F} quantified the migration efficiency in a model parametrizing the stellar migration distance as a Gaussian probability distribution broadening with time.
By fitting the model to the age-metallicity distribution of red clump stars, they find the Gaussian to have a standard deviation of $3.6\pm 0.1$~kpc at 8~Gyr.
This indicates that stellar migration in the disc is efficient; however, whether this effect can cause a dip in the SFH like the one found in this study remains to be investigated.

\section{Discussion} \label{sec:discussion}
Overall, the synthetic tests of the method are successful.
The method works as intended on fairly modest sample sizes with realistic parameter uncertainties.
The recovery of the true age-metallicity distribution is never perfect, but it is almost always more accurate than the distribution of individual age estimates.
In the following we give a brief discussion of specific advantages and disadvantages of the method as well as further challenges introduced when applying the method to real data.

\subsection{Advantages and disadvantages of using the SAMD/SAD}
One of the strong points of this method is that no stars need to be excluded from the analysis based on the precision of their age estimate, as long as it is not significantly biased.
In practice this means that one can make a fairly simple selection of stars on the turnoff, and the SAMD will be informed by the individual $\mathcal{G}$~functions of the entire sample -- even those that provide little to no constraints on the individual parameters.
When using individual ages, the ones with the largest uncertainties are often excluded from the analysis, lowering the effective sample size.
In the original formulation of the method by \citet{1999MNRAS.304..705H} the tests were in fact carried out without excluding giants from the sample.
However, their samples had much fewer giants than the ones we have tested, and it is likely that their results would have suffered if the number of giants was increased.

In practically all cases tested here, the SAD gives a better indication of the true age distribution than the histogram of individual age estimates.
This is true even when restricting the comparison to those stars with relative age uncertainties lower than 25\%.
The SAD therefore performs better in recovering the relative fraction of stars at different ages as seen e.g. in \autoref{fig:ages_1D_comparison}.
One also avoids having to choose a statistic for the individual age estimates.

In this work we have focused on using the method on the two-dimensional age-metallicity PDFs in order to estimate the SAMD.
Then the SAD can be obtained by marginalizing the metallicity.
But the SAD can also be obtained directly by applying the method on the one-dimensional age PDFs.
One of the advantages of working with the SAMD is, as demonstrated by the GCS example, that it gives a different impression of the SFH when the joint age-metallicity distribution is computed.
In the SAMD one sees that the dip in the SFH at an age of 10~Gyr is due to what appears to be a separate older and more metal-poor pouplation.
This is obtained without having to divide the sample into bins according to their individual metallicities and the metallicity uncertainties, as well as the age-metallicity correlations, are automatically taken into account in the solution.

Another feature of the method, which has not been explored in this work, is that it is not restricted to use with individual PDFs obtained from isochrone fitting.
The description of the method in section \ref{sec:method_sample_dist} and Appendix \ref{sec:appA} is quite general and does not depend on the source of the $\mathcal{G}$~functions.
Thus, it can be applied with individual age PDFs from other methods such as gyrochronology \citep{2007ApJ...669.1167B}.

Finally, there is only a single parameter to tune in the method: the regularization parameter $\alpha$.
The solution is not parametrized in terms of other functions and it can in principle take any form informed by the data and restricted only by $\alpha$.
However, this also leads to one of the downsides of the method, namely that the result depends on the chosen value of $\alpha$, and the optimal value depends on the problem.
Although we describe how one can choose this parameter in Appendix \ref{sec:appB}, there is no guarantee that this is the optimal way to choose it for every application of the method.
Hence, one must be cautious with the application of the method and the interpretation of its results if it is difficult to choose the optimal value of $\alpha$.

It is worth mentioning that the SAMD will of course be biased if the individual $\mathcal{G}$~functions are biased.
This is most obvious when including giant stars without the luxury of seismic parameters.
It can also happen near the turnoff, when the parameter uncertainties are large, simply due to the nonlinear transformation between surface parameters and age.
This is what \citet{2019A&A...629A.127M} try to work around by inferring the SFH using the estimated age distributions of synthetic mono-age populations.
By assuming that the bias in the sample age distribution and the mono-age distributions is the same, this bias is reduced.
This is an interesting approach and one that should be compared with the method presented in this work in the future.

A final disadvantage is that there is no uncertainty estimate built into the method.
Even if an uncertainty of the distribution is estimated e.g. by resampling, there is additional uncertainty associated with the choice of $\alpha$ which is not straightforward to quantify.

\subsection{Practical challenges with real samples}
Moving to real observations introduces additional challenges and potential biases in the results.
These challenges are not strictly related to the estimation of the SAMD.
Instead, the challenge lies in properly estimating the individual $\mathcal{G}$~functions.
Since we have only tested the method on $\mathcal{G}$~functions from isochrone fitting, it is relevant to discuss some of the challenges one might face in this case.

First of all, neither the models nor the data are perfect which limits the accuracy of the estimated age distribution.
For example, the value of $T_{\mathrm{eff}}$ can differ by up to $200$~K depending on the calibration \citep{2010A&A...512A..54C} which may significantly change the age estimate.
Additionally, stellar models calculated with different input physics will have different temperatures and luminosities for the same age and metallicity.
This is seen most clearly as a relative shift in the position of the turnoff \citep[see e.g.][]{2019MNRAS.482..895S}
These differences mainly impact the absolute age scale while the relative ages are less affected.
Taking the SAMD for the GCS sample as an example, this means that the minimum observed at an age of 10~Gyr may shift slightly up or down in age if a different set of models was used; however, the minimum would likely remain with a similar width.

Other challenges exist even if the models and observations were perfect.
A real sample may include unresolved binaries which are assumed to be single stars at a higher luminosity.
\citet{1999MNRAS.304..705H} tested the impact of this on the SFH by creating a synthetic sample with a Gaussian age distribution centred on 7.5~Gyr and a binary fraction of 0.5.
When attempting to recover the SFH assuming a binary fraction of 0, the binaries show up as a small old tail extending up to 5~Gyr above the edge of the true SFH.
\citet{2013MNRAS.428..763S} carried out a similar test with a binary fraction of 0.4 and also found a bias towards higher ages in their age-metallicity distributions.
Blue stragglers, stars that have gained mass through binary evolution, also introduce a bias by appearing younger than they really are \citep[e.g.,][]{2018MNRAS.481.5062B}.
In a sample like the GCS which is dominated by young stars this is not a major issue since the blue stragglers only make up a small fraction of the real sample.
However, in older populations they will stand out and look like a young component which does not exist in reality.

\subsection{Applying the method in practice}
In order to facilitate application of the method, we make available a Python implementation of the algorithm, as laid out in Appendix~\ref{sec:appA}, along with an example script for running it on the Milky Way-like turnoff sample from section~\ref{sec:MW_pop} \citep{samd_2020}.\footnote{The example implementation can be found here: \url{https://github.com/csahlholdt/SAMD}.}
When applying this algorithm to other samples, the method should initially be run for a wide range of $\alpha$ values.
This range should cover several orders of magnitude, e.g. from $\alpha = 1$ to $\alpha = 10^7$.
From there, the optimal range of $\alpha$ can be narrowed down following the recommendations given in Appendix~\ref{sec:appB}, or by any other method which can be shown to work.

All the tests carried out in this work were made on samples with known parameter distributions (except for the GCS example); therefore, the choice of $\alpha$ could in principle be tuned to achieve the best possible recovery of this distribution.
This shows that a value of $\alpha$ usually exists for which the SAMD is a good representation of the true distribution; however, it is not necessarily straightforward to determine this value.
Therefore, in real applications, one must be careful not to over-interpret every little wrinkle in the distribution if there is considerable uncertainty in the choice of the value of $\alpha$.
For example, if the chosen value of $\alpha$ is too low, short scale oscillations show up in the SAMD even when the true distribution is completely smooth on small scales (see \autoref{fig:alpha_progression}).
The most robust conclusions are those that depend only weakly on the choice of $\alpha$.

Ideally, if one wants to know if a certain parameter distribution is robust, a number of synthetic samples should be analysed in the same way as the real data.
For example, one sample with a completely flat parameter distribution and another with the distribution recovered from the real data.
If these distributions cannot be distinguished with this method then the distribution recovered from the real data should not be trusted.
Such synthetic experiments can also provide information about the best choice of $\alpha$ specific to the sample at hand.

\section{Conclusions} \label{sec:conclusion}
We have implemented a new algorithm for estimating the distribution $\phi(\theta)$, of a set of parameters $\theta$, from a sample of individual PDFs $\mathcal{G}_{i}(\theta|\mathbf{x})$.
The method has been applied to the problem of estimating the age-metallicity distribution of a sample of stars (the SAMD) with $\mathcal{G}_{i}(\theta|\mathbf{x})$ estimated by isochrone fitting with $\mathbf{x} = (T_{\mathrm{eff}}, [\mathrm{Fe}/\mathrm{H}], \mathrm{K_s}, \varpi)$, which are, respectively, the effective temperature, metallicity, apparent magnitude in the 2MASS $\mathrm{K_s}$~band, and the parallax.
The algorithm has only a single free parameter which is used to set the degree of smoothness of the solution.
Aside from this constraint, the shape of the solution is entirely determined by the individual age-metallicity PDFs.

The method is shown to work as intended for synthetic mono-age populations drawn from a single isochrone.
For realistic parameter uncertainties, a sample size of about 100 is sufficient to accurately recover the true age of a population of turnoff stars to within 0.2~Gyr.
A larger sample size and smaller parameter uncertainties increase the precision of the recovered age as measured by the width of the SAMD.
This is not the case for distributions of individually estimated ages which show the same width for different sample sizes.
Additionally, mono-age populations with different ages and metallicities stand out more clearly in the SAMD than they do in the space of individually estimated ages and metallicities.

Using turnoff stars only, we are also able to recover age-metallicity and age distributions for synthetic samples with a number of different extended age-metallicity distributions, e.g. a flat distribution, a double guassian distribution, and one with the age-metallicity relation taken from the Besan\c{c}on Galaxy model.
Modest sample sizes of 1000 stars (or about 100 per Gyr), as tested here, are found to be sufficient, but larger sample sizes increase the accuracy of the solution.
By including asteroseismic parameters in the isochrone fitting, it is also possible to recover the parameter distribution using giant stars only.

The age distributions recovered using the SAMD are not perfect, especially at ages greater than about 11~Gyr, but they are always closer to the true distribution than the histogram of individual age estimates.
This is also true when only individual ages with relative uncertainties below 25 or 15 per cent are included.
By making this selection, the age distribution is usually biased towards a higher fraction of younger stars since these have more precise age estimates.

As a first application to a real sample of stars, we have re-derived the age-metallicity relation of the Geneva-Copenhagen survey.
For this sample the main features of the SAMD are consistent with previous studies, but it also indicates that there is a minimum in the distribution of stars at an age of 10~Gyr.
Such a minimum is not obvious in the distribution of individual ages and metallicities.
This may reflect a real minimum in the SFH, separating two stellar populations with different origins.

This method will be useful in studies of Galactic stellar populations combining \textit{Gaia} astrometry with spectroscopic surveys and, potentially, asteroseismic data.
It can give a more precise estimate of the age-metallicity relation than the distribution of individually estimated ages and thereby help put stronger constraints on the order of formation of different stellar populations.
For example, it can be used to estimate and compare the age-metallicity relations of stellar populations with different kinematic properties, e.g. in-situ and accreted stars.
Such results could ultimately be compared with those of hydrodynamic Milky Way simulations to gain more insight into the formation history of the Galaxy.
It also has the potential to be applied with individual age PDFs from other methods than isochrone fitting, e.g. gyrochronology.
However, as with all stellar age estimates, it must be applied and interpreted with care in cases where biases which are not explored in this work may be introduced.
This includes age biases due to inaccuracies in the stellar models, uncertain or biased stellar parameters ($T_{\mathrm{eff}}$, $[\mathrm{Fe}/\mathrm{H}]$, etc.), and the presence of unresolved binaries or blue stragglers.
These issues all affect the individual age PDFs and thus naturally propagate into the SAMD.

\section*{Acknowledgements}
CLS was supported by the grant 2016-03412 from the Swedish Research Council.
LL acknowledges support from the Swedish National Space Board.
This research has made use of the SIMBAD database,
operated at CDS, Strasbourg, France.
This work has made use of data from the European Space Agency (ESA) mission
{\it Gaia} (\url{https://www.cosmos.esa.int/gaia}), processed by the {\it Gaia}
Data Processing and Analysis Consortium (DPAC,
\url{https://www.cosmos.esa.int/web/gaia/dpac/consortium}). Funding for the DPAC
has been provided by national institutions, in particular the institutions
participating in the {\it Gaia} Multilateral Agreement.

\section*{Data availability}
Parts of the data underlying this article, along with an implementation of the algorithm, are available in Zenodo, at \url{https://dx.doi.org/10.5281/zenodo.3941733}.
The synthetic data underlying this article will be shared on reasonable request to the corresponding author.

%%%%%%%%%%%%%%%%%%%%%%%%%%%%%%%%%%%%%%%%%%%%%%%%%%



\bibliographystyle{mnras}
\bibliography{samd} % if your bibtex file is called example.bib

%\begin{thebibliography}{99}
%\end{thebibliography}

%%%%%%%%%%%%%%%%%%%%%%%%%%%%%%%%%%%%%%%%%%%%%%%%%%

%%%%%%%%%%%%%%%%% APPENDICES %%%%%%%%%%%%%%%%%%%%%

\appendix

\section{Discretization and minimization} \label{sec:appA}
Here we describe the implementation of the method to estimate $\hat\phi$ by minimizing equation \eqref{eq:lnL_final}.
To facilitate this minimization, an additional term, $p(\phi)$, is introduced.
This term can be thought of as a prior on $\phi$, and we choose the function
\begin{equation} \label{eq:phi_prior}
    p(\phi) = \exp\left( -\frac{1}{\beta} \int\phi(\theta) \ln\left( \frac{\phi(\theta)}{\Phi(\theta)} \right)\mathrm{d}\theta \right) \, ,
\end{equation}
which is known as the relative entropy of $\phi$ with respect to $\Phi$.
This function has a number of desirable properties.
Firstly, it is maximized for $\phi = \Phi$, with $\Phi$ being a known non-negative function, satisfying equation \eqref{eq:norm_req}.
We choose $\Phi$ to be a flat (constant) density.
Secondly, $p(\phi)$ depends on the parameter $\beta \geq 0$ which can be varied at will, and which regulates how strongly $p$ favours $\Phi$.
A value of $\beta = 0$ corresponds to an infinitely strong prior which forces the solution $\phi = \Phi$.
Finally, $p(\phi)$ incorporates the non-negativity of $\phi$ by assigning zero probability to functions $\phi$ that are negative for some $\theta$.
By defining $P(\phi) = L'(\phi)p(\phi)$, the problem becomes to minimize the expression
\begin{align} \label{eq:lnP_final}
    \begin{split}
        -\ln P ={}& \frac{1}{\beta}\int\phi(\theta) \ln\left( \frac{\phi(\theta)}{\Phi(\theta)} \right)\mathrm{d}\theta - \sum_{i} \ln\left( \int \mathcal{G}_{i}(\theta)\phi(\theta)\mathrm{d}\theta \right)\\ {}&+ \alpha \int\left(\sum_{q=1}^{N_{\mathrm{dim}}}s_{q}^{2}\frac{\partial^2\phi}{\partial\theta_{q}^{2}} \right)^2\mathrm{d}\theta \, ,
    \end{split}
\end{align}
or, equivalently, to minimize $J = -\beta \ln P$
\begin{align} \label{eq:J_final}
    \begin{split}
        J(\phi) ={}& \int\phi(\theta) \ln\left( \frac{\phi(\theta)}{\Phi(\theta)} \right)\mathrm{d}\theta - \beta\sum_{i} \ln\left( \int \mathcal{G}_{i}(\theta)\phi(\theta)\mathrm{d}\theta \right) \\
        {}&+ \beta \alpha \int\left(\sum_{q=1}^{N_{\mathrm{dim}}}s_{q}^{2}\frac{\partial^2\phi}{\partial\theta_{q}^{2}} \right)^2\mathrm{d}\theta \, .
    \end{split}
\end{align}
For $\beta\rightarrow \infty$ this is equivalent to minimizing equation \eqref{eq:lnL_final}.
The idea is to start at $\beta = 0$, where the solution is known, and then increase $\beta$ gradually so the solution changes slowly enough for proper convergence.
How this works is seen more clearly at the end of this section when the details of the implementation have been described.

In practice, the algorithm works with discrete variables, so $\theta$ only takes a finite set of values $\theta_{j}$, $j = 0\dots m-1$, where each index $j$ corresponds to a specific set of parameters.
For example, for the age-metallicity distribution, each $j$ corresponds to a combination of $\tau$ and $\zeta$, and there is $m$ such combinations.
Similarly, $\mathcal{G}$ is evaluated at the discrete points $\mathcal{G}_{ij} = \mathcal{G}_{i}(\theta_{j})$.
We also use the short-hand notation $\phi_{j} = \phi(\theta_{j})$.

Equation \eqref{eq:norm_req} can now be written
\begin{equation} \label{eq:norm_req_discrete}
    \sum_{j}w_{j}\phi_{j} = 1 \, ,
\end{equation}
where $w_{j}$ are suitably defined weights.
We set all weights equal to the same value $w_{j} = 1/m$.
The integrals in equation \eqref{eq:phi_likelihood} can be written
\begin{equation} \label{eq:ui}
    u_{i} \equiv \int \mathcal{G}_{i}(\theta) \phi(\theta)\mathrm{d}\theta = \sum_{j} \mathcal{G}_{ij}w_{j}\phi_{j} \, ,
\end{equation}
and the sum of derivatives in the final term of equation \eqref{eq:J_final} can be written
\begin{equation} \label{eq:vk}
    v_{k} \equiv \left(\sum_{q=1}^{N_{\mathrm{dim}}}s_{q}^{2}\frac{\partial^2\phi}{\partial\theta_{q}^{2}} \right)_{\theta = \theta_{k}} = \sum_{j} \mathcal{T}_{kj}w_{j}\phi_{j} \, , \; \;\;\; k = 0\dots m-1 \, ,
\end{equation}
where $\mathcal{T}$ is the appropriate matrix representation of the second derivative operator which depends on the number of dimensions, $N_{\mathrm{dim}}$, of $\theta$.
The factor $s_{q}$ has been absorbed into $\mathcal{T}$.
In one dimension, we use the lowest order finite difference coefficients for the second derivative which gives the approximation
\begin{equation}
    \left(\frac{\mathrm{d}^2\phi}{\mathrm{d}\theta^{2}}\right)_{\theta = \theta_{k}} \approx \frac{1}{\Delta\theta^{2}} \left[\phi(\theta_{k-1})-2\phi(\theta_{k})+\phi(\theta_{k+1})\right] \, ,
\end{equation}
where $\Delta\theta = \theta_{j+1}-\theta_{j}$ is the grid spacing.
By choosing the scale $s = \Delta\theta$, this factor cancels and $\mathcal{T}$ is the $m$-by-$m$ matrix with entries
\begin{equation}
    \mathcal{T}_{\mathrm{1D},kj} =
    \begin{cases}
        -2 \, &\text{if  } k=j \, , \\
        1 \, &\text{if  } k=j\pm1 \, , \\
        0 \, &\text{otherwise} \, .
    \end{cases}
\end{equation}
In two dimensions, which is the highest we will consider here, $\mathcal{T}$ is also an $m$-by-$m$ matrix; however, its layout is different and depends on the mapping from 2D space to the linear index $j$.
Say, for example, the two parameters are $\theta=(\tau, \zeta)$ with discrete values $\tau_{r}$, $r = 0\dots R-1$, and $\zeta_{s}$, $s = 0\dots S-1$.
Then the approximate second order derivative is
\begin{align} \label{eq:2D_derivative1}
    \begin{split}
        \left(\frac{\partial^2\phi}{\partial\tau^{2}}+\frac{\partial^2\phi}{\partial\zeta^{2}}\right)_{(\tau, \zeta) = (\tau_{r}, \zeta_{s})} {}&\approx \frac{\phi(\tau_{r-1},\zeta_{s})-2\phi(\tau_{r},\zeta_{s})+\phi(\tau_{r+1},\zeta_{s})}{\Delta\tau^{2}} \\
        {}&+  \frac{\phi(\tau_{r},\zeta_{s-1})-2\phi(\tau_{r},\zeta_{s})+\phi(\tau_{r},\zeta_{s+1})}{\Delta\zeta^{2}}\, .
    \end{split}
\end{align}
Defining the one-dimensional index as
\begin{equation} \label{eq:1d_index}
    k=sR+r \, ,
\end{equation}
the right-hand side of equation \eqref{eq:2D_derivative1} can be written
\begin{equation}
    \frac{\phi_{k-1}-2\phi_{k}+\phi_{k+1}}{\Delta\tau} + \frac{\phi_{k-R}-2\phi_{k}+\phi_{k+R}}{\Delta\zeta} \, .
\end{equation}
This can be represented by an $S$-by-$S$ block matrix with each block being an $R$-by-$R$ matrix.
Like in one dimension, this is an $m$-by-$m$ matrix since $S\times R = m$.
In this case we also set the scale factors equal to the grid spacings.
Then each $R$-by-$R$ block on the main diagonal of $\mathcal{T}$ is given by
\begin{equation}
    \mathcal{T}_{\mathrm{2D(diag)},rr^{\prime}} =
    \begin{cases}
        -4 \, &\text{if  } r=r^{\prime} \, , \\
        1 \, &\text{if  } r=r^{\prime}\pm1 \, , \\
        0 \, &\text{otherwise} \, ,
    \end{cases}
\end{equation}
and each super- and subdiagonal block is the identity matrix
\begin{equation}
    \mathcal{T}_{\mathrm{2D(subdiag)},rr^{\prime}} =
    \begin{cases}
        1 \, &\text{if  } r=r^{\prime} \, , \\
        0 \, &\text{otherwise} \, ,
    \end{cases}
\end{equation}
and all remaining blocks are $0$.
Of course, in this case the parameter index $j$ must have the same mapping to the two-dimensional space as $k$ given in equation \eqref{eq:1d_index}.

Using the definitions given in equations \eqref{eq:ui} and \eqref{eq:vk} the discrete version of equation \eqref{eq:J_final} becomes
\begin{equation}
    J(\phi) = \sum_{j} w_{j}\phi_{j} \ln\left( \frac{\phi_{j}}{\Phi_{j}} \right) - \beta \sum_{i} \ln u_{i} + \beta\alpha \sum_{k} v_{k}^{2} \, .
\end{equation}
To incorporate the constraint of equation \eqref{eq:norm_req_discrete}, we add a Lagrange multiplier $\lambda$, so that
\begin{equation}
    K(\phi, \lambda) = \sum_{j} w_{j}\phi_{j} \ln\left( \frac{\phi_{j}}{\Phi_{j}} \right) - \beta \sum_{i} \ln u_{i} + \beta\alpha \sum_{k} v_{k}^{2} + \lambda \left( \sum_{j} w_{j}\phi_{j} - 1 \right) \, ,
\end{equation}
should be minimized with respect to the $m+1$ arguments $\phi_{j}$ and $\lambda$.

The minimization is done by setting the partial derivatives of $K$ to zero and solving the resulting non-linear system of equations by Newton-Raphson iteration.
Remembering that $u_{i}$ and $v_{k}$ depends on all the $\phi_{j}$ through equations \eqref{eq:ui} and \eqref{eq:vk}, the partial derivatives are readily obtained:
\begin{align} \label{eq:residual1}
    \begin{split}
    r_{j} \equiv \frac{\partial K}{\partial \phi_{j}}
           ={}& w_{j}\left[1 + \ln\left( \frac{\phi_{j}}{\Phi_{j}} \right) \right] -
            \beta \sum_{i} \mathcal{G}_{ij} w_{j} u_{i}^{-1} \\
            {}&+2\alpha\beta\sum_{k} \mathcal{T}_{kj}w_{j}v_{k} +
            \lambda w_{j} \, ,
    \end{split}
\end{align}
\begin{equation}
    R \equiv \frac{\partial K}{\partial \lambda} = \sum_{j}w_{j}\phi_{j} - 1 \, . 
\end{equation}
The residuals are non-zero as long as the correct solution has not been found.
To improve the solution, we look for corrections $\Delta\phi_{j}$ and $\Delta\lambda$ that will make the residuals smaller.
The predicted residuals after correction, obtained by the linear expansion, are
\begin{align}
    r_{j}^{(\mathrm{new})} = r_{j}^{(\mathrm{old})} + \sum_{l} \frac{\partial r_{j}}{\partial \phi_{l}} \Delta\phi_{j} + \frac{\partial r_{j}}{\partial \lambda} \Delta\lambda \, ,\\
    R^{(\mathrm{new})} = R^{(\mathrm{old})} + \sum_{l} \frac{\partial R}{\partial \phi_{l}} \Delta\phi_{j} + \frac{\partial R}{\partial \lambda} \Delta\lambda \, .
\end{align}
Setting these to zero gives a set of linear equations for the corrections:
\begin{equation} \label{eq:corr_matrix}
    \begin{bmatrix}
        \boldsymbol{H} & \boldsymbol{w} \\
        \boldsymbol{w} & 0
    \end{bmatrix}
    \begin{bmatrix}
        \boldsymbol{\Delta\phi} \\
        \Delta\lambda
    \end{bmatrix}
    =
    \begin{bmatrix}
        \boldsymbol{-r} \\
        -R
    \end{bmatrix} \, .
\end{equation}
Here $\boldsymbol{H}$ is the Hessian matrix with elements
\begin{align}
    \begin{split}
        H_{jl} = \frac{\partial^{2}K}{\partial\phi_{j}\partial\phi_{l}} ={}& \delta_{jl}w_{j}\phi_{j}^{-1} + \beta\sum_{i}\mathcal{G}_{ij}w_{j}\mathcal{G}_{il}w_{l}u_{i}^{-2}\\
        {}&+ 2\alpha\beta\sum_{k}\mathcal{T}_{kj}w_{j}\mathcal{T}_{kl}w_{l} \, ,
    \end{split}
\end{align}
where $\delta_{jl}$ is the Kronecker delta and $\boldsymbol{w}$, $\boldsymbol{r}$, and $\boldsymbol{\Delta\phi}$ are $m\times 1$ vectors with elements $w_{j}$, $r_{j}$, and $\Delta\phi_{j}$.
The last row and column in equation \eqref{eq:corr_matrix} follow from $\partial R/\partial\phi_{j}=w_{j}$ and $\partial R/\partial\lambda=0$.

When $K$ is minimized by Newton-Raphson iteration, it may happen that $\phi_{j}+\Delta\phi_{j}$ becomes negative for some $j$.
This simply means that the corrections are too large in relation to the range in which the linear expansion is valid.
It is then necessary to take a smaller step.
To ensure that $\phi_{j}>0$, we apply the correction as
\begin{equation}
    \phi_{j}\leftarrow \phi_{j} + f\Delta\phi_{j} \, , \;\;\; \lambda\leftarrow \lambda + f\Delta\lambda \, ,
\end{equation}
where $0<f\leq1$ is chosen to have $\mathrm{min}(\phi_{j})>0$ after the correction.

We can now find the solution in the special case of $\beta=0$ where the minimizing function is $\phi=\Phi$.
Inserting this in equation \eqref{eq:residual1} and solving $r_j=0$ shows that in this case $\lambda=-1$.
Now if we make $\beta$ only slightly greater than $0$, we can assume that the solution is only slightly different, and can be obtained by relatively few Newton-Raphson iterations.
We can then increase $\beta$ in small steps, with a moderate number of iterations in each step, until further increase only leads to negligible change of the solution.
In this case $\beta$ acts merely as a stabilising parameter during minimization which ensures gradual convergence from the initial function $\Phi$ towards a solution defined by the minimization of equation \ref{eq:lnL_final}.
This solution still depends on the choice of $\alpha$ which is explored in appendix \ref{sec:appB}.

\section{Choosing $\alpha$} \label{sec:appB}
\begin{figure*}
\centering
\includegraphics[width=\textwidth]{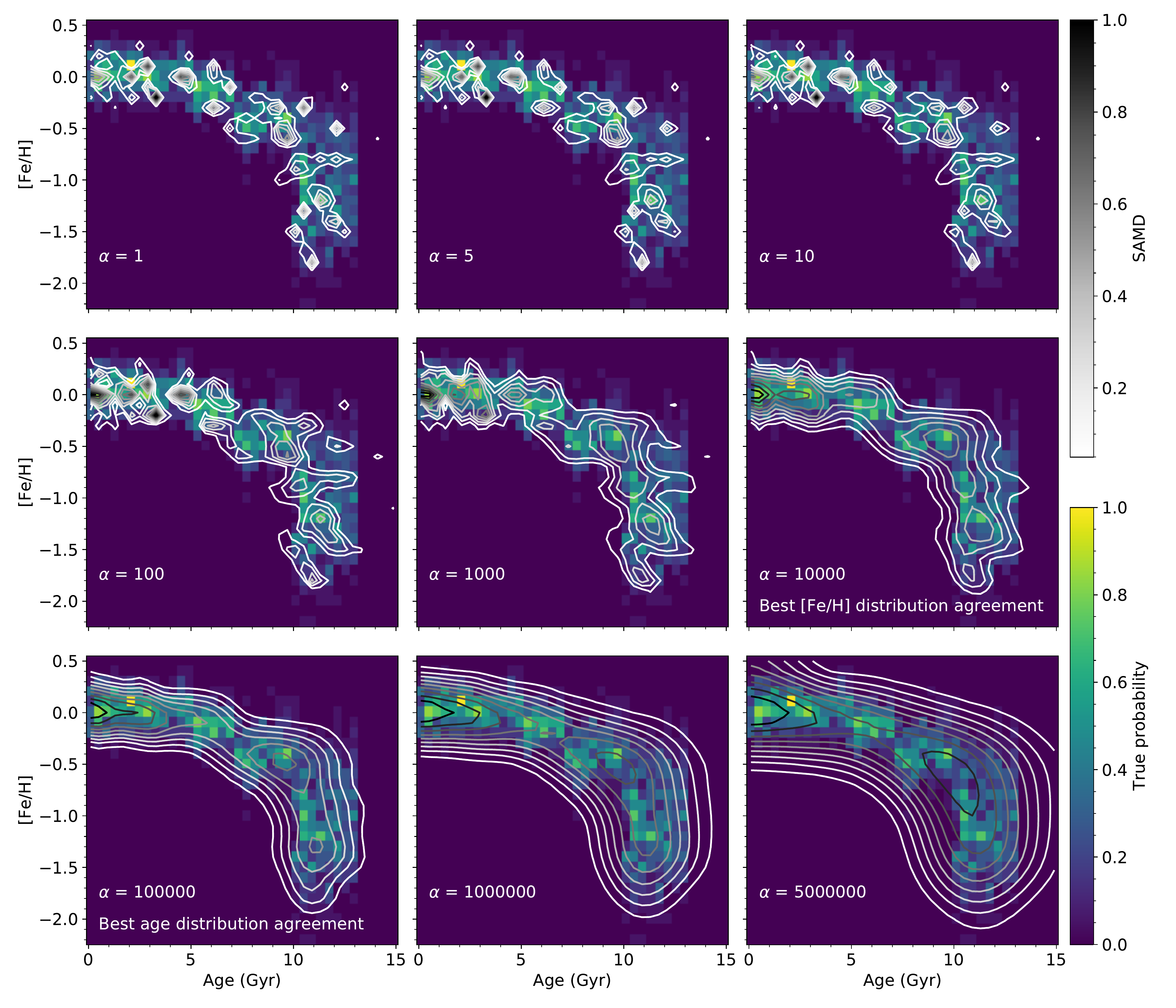}
\caption{SAMD contours, for a range of $\alpha$ values as annotated, on top of a 2D histogram of the true age-metallicity distribution of the Milky Way-like turnoff sample (see section~\ref{sec:MW_pop}).}
\label{fig:alpha_progression}
\end{figure*}

The choice of the regularization parameter is important for samples with extended parameter distributions.
\autoref{fig:alpha_progression} shows an example of how the solution changes as one increases $\alpha$ from a value of $1$, which has almost no impact on the solution, to a high value resulting in over-smoothing.
With no smoothing, the solution is made up of a number of very narrow spikes.
These spikes do more or less trace out the true distribution we are trying to recover, but the true distribution is smooth.
In this case a value of $\alpha$ somewhere around $10^5$--$10^6$ seems optimal and anything higher leads to a solution which is too smooth.
However, different problems will have different optimal values which may be significantly lower or higher.

\begin{figure*}
\centering
\includegraphics[width=\textwidth]{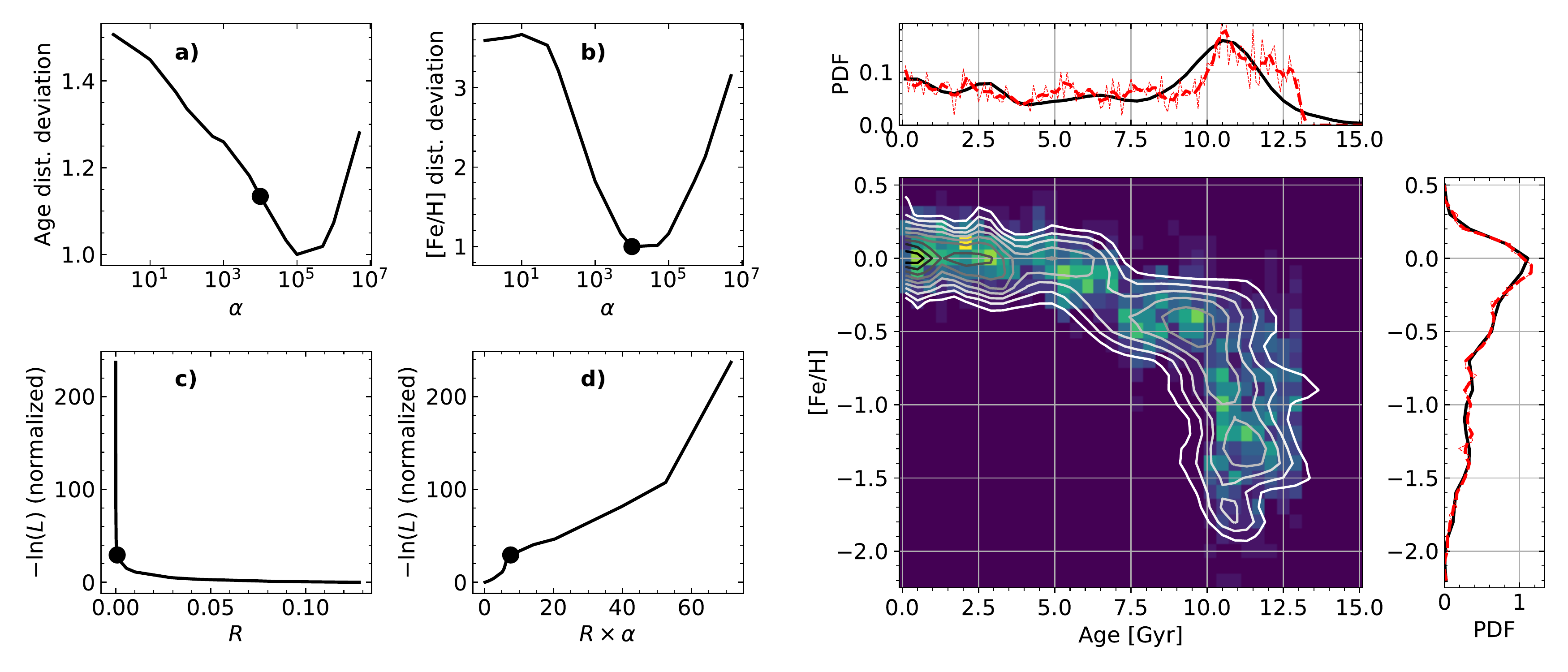}
\caption{Right-hand side: SAMD contours on top of a 2D histogram of the true age-metallicity distribution of the Milky Way-like turnoff sample (see section~\ref{sec:MW_pop}).
The two adjacent panels show the age and metallicity distributions of the marginalized SAMD as black lines.
The corresponding true distributions are shown as thin red dashed lines with smoothed versions shown as thick red dashed lines.
Left-hand side: a) The sum of squared differences between the true age distribution and the SAD as a function of $\alpha$, normalized to a minimum value of 1.
b) The same as a) but for the metallicity distributions.
c) The trade-off curve of negative log-likelihood against smoothness parameter for the same range of $\alpha$ as shown in panels a) and b).
The negative log-likelihood has been normalized to a value of 0 at $\alpha=0$ by subtracting the value at $\alpha=0$.
d) Like c) but with the smoothness parameter multiplied by $\alpha$.
In all panels the black dot indicates the solution shown on the right-hand side.} 
\label{fig:alpha_multi_age_to}
\end{figure*}

\begin{figure*}
\centering
\includegraphics[width=\textwidth]{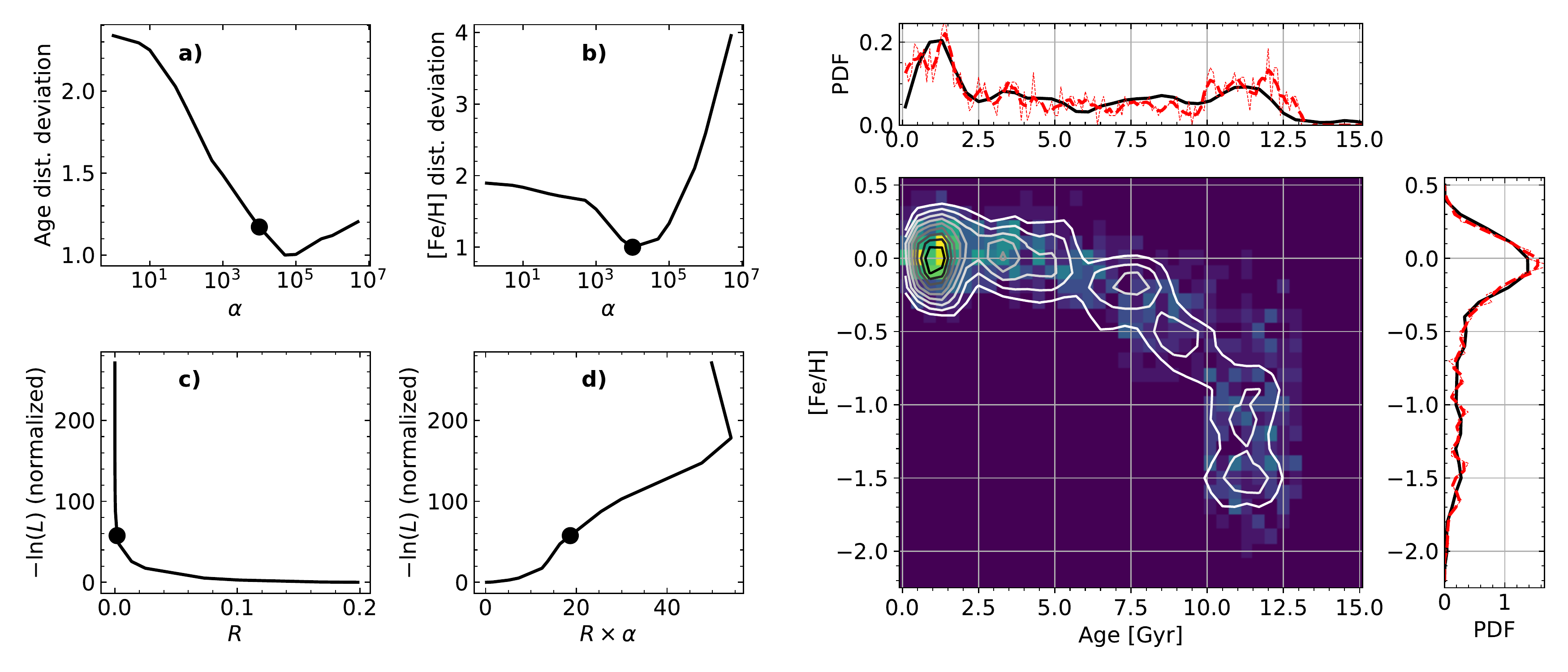}
\caption{Like \autoref{fig:alpha_multi_age_to} but for the Milky Way-like giant sample using seismic parameters in the isochrone fits (see section~\ref{sec:seismic_info}).}
\label{fig:alpha_multi_age_giants}
\end{figure*}

\begin{figure*}
\centering
\includegraphics[width=\textwidth]{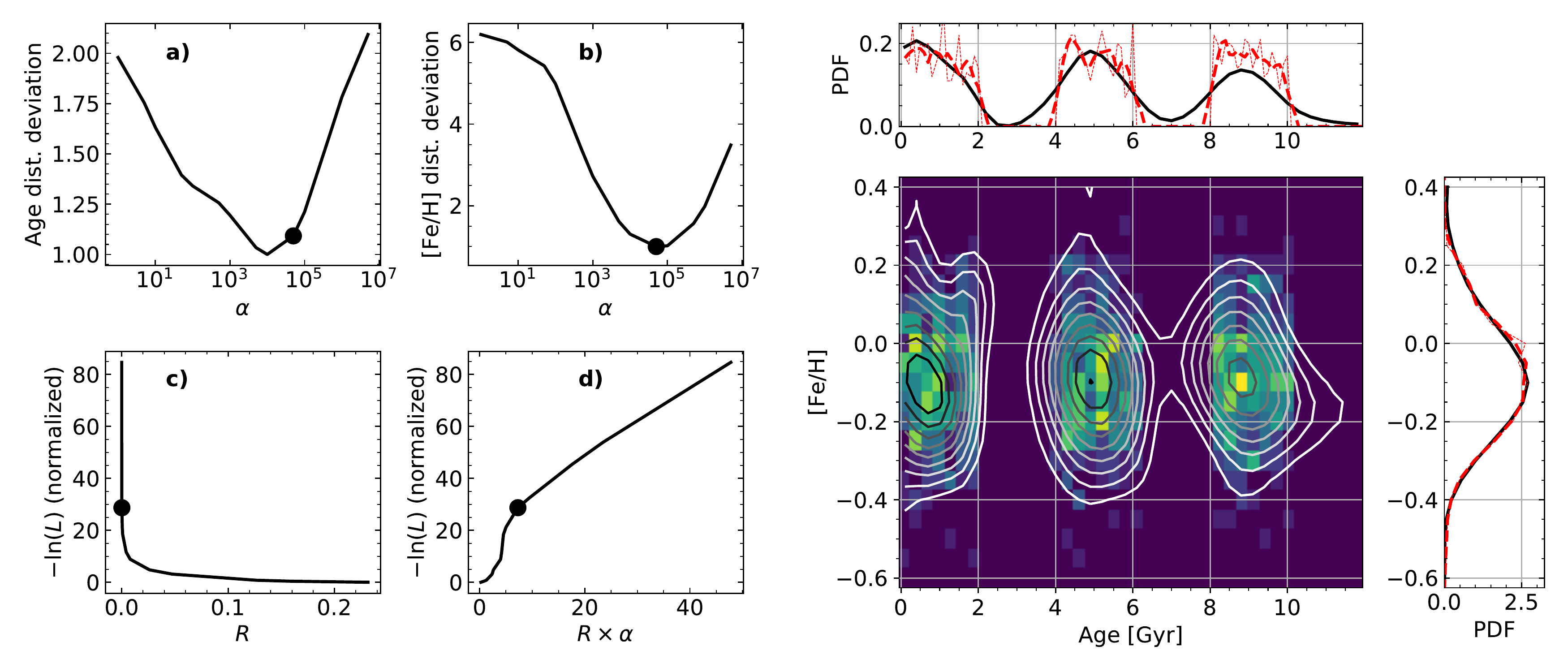}
\caption{Like \autoref{fig:alpha_multi_age_to} but for the on-off age distribution described in section~\ref{sec:extended_age_pops}.}
\label{fig:alpha_on_off}
\end{figure*}

\begin{figure*}
\centering
\includegraphics[width=\textwidth]{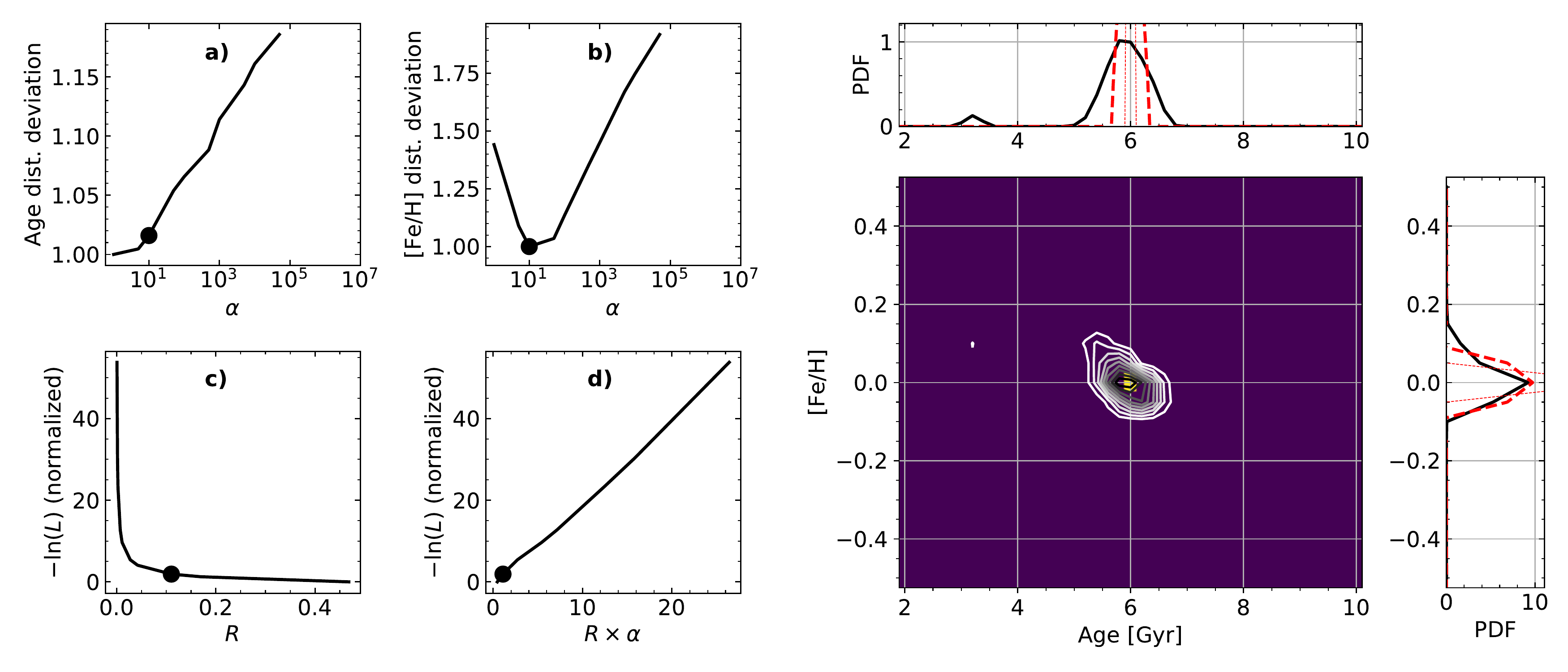}
\caption{Like \autoref{fig:alpha_multi_age_to} but for the mono-age population of 1000 turnoff stars with an age of 6~Gyr.}
\label{fig:alpha_single_iso}
\end{figure*}

Here we describe how one can choose the value of the regularization parameter $\alpha$ in the case of age-metallicity or age distributions.
The optimal choice of $\alpha$ is simple for synthetic data: In this case we choose the value which gives the best agreement between the SAMD and the true age-metallicity distribution.
However, when analysing real data the true distribution is unknown, and it is necessary to have other criteria for choosing $\alpha$.
By looking at the optimal $\alpha$ for synthetic samples, where the age distributions are known, we may identify such criteria.
It is important to stress that no single criterion is perfect or universally applicable.
Therefore, the criteria presented here should be taken as recommendations only, and one should always critically asses the choice for each specific application.

Since we only consider the case where the individual stellar metallicities are known, and used as input for the isochrone fits, the optimal SAMD is expected to recover the input metallicity distribution very well.
Therefore, we test whether a good agreement between the observed and recovered metallicity distributions is a good criterion for choosing $\alpha$.
In \autoref{fig:alpha_multi_age_to} the SAMD for the ``Milky Way-like" turnoff population is shown with $\alpha$ chosen such that the difference between the true (input) and recovered (SAMD) metallicity distributions is minimized.
As expected, the distributions are nearly identical.
Additionally, this choice of $\alpha$ turns out to be very close to the value which gives the optimal agreement between the true and recovered age distributions.
This result is reproduced for other samples (Figures \ref{fig:alpha_multi_age_giants} and \ref{fig:alpha_on_off}) indicating that the recovery of the known metallicity distribution may be a good criterion for choosing $\alpha$.

One can also compare the two terms in equation \eqref{eq:lnL_final} for the optimal solution.
Equation \eqref{eq:lnL_final} can be written $-\ln L' = -\ln L + \alpha R$, where $-\ln L$, given by equation \eqref{eq:lnL}, is the likelihood which measures the agreement between the SAMD and the data (the individual $\mathcal{G}$~functions), and $R$ is the integral which measures the curvature of the SAMD and is equal to $0$ for a flat distribution.
Plotting $-\ln L$ against $R$ (also shown in \autoref{fig:alpha_multi_age_to}) results in a trade-off curve which is a common feature in inversion problems.
Lower $-\ln L$ means better agreement with the data (obtained with a low value of $\alpha$) while a lower $R$ means a smoother solution (forced by increasing $\alpha$) .
In the examples shown here, the optimal solution is always near the bend of the curve.
However, the location of this bend depends on the range of $\alpha$-values tested because this changes the scales of the axes.
It may be better to just consider the value of $-\ln L$ since this is the quantity which carries statistical information about the solution.
The normalized value ($-\ln L(\alpha_{\mathrm{best}})-(-\ln L(\alpha = 0))$) is in the range 30--60 for the samples shown in Figures \ref{fig:alpha_multi_age_to} to \ref{fig:alpha_on_off}.
This indicates that the optimal solution is likely to be found in the range of $-\ln L$ about 10--100 above the value at $\alpha=0$.
However, this will surely vary within a larger interval when considering more age-metallicity distributions and sample sizes.

One can also consider the curve obtained by plotting $-\ln L$ against $R\times\alpha$.
This curve shows a steep transition between two flatter sections, and the optimal solution appear to always fall shortly after the steepest part.
It is unclear what exactly causes this shape of the curve, but it may be that the steepest part marks a transition from a distribution with narrow spikes to one that is varying smoothly everywhere.
In the example shown in \autoref{fig:alpha_progression}, the steepest part is around $\alpha = 1000$ where the solution transitions from a number of ``probability islands" at $\alpha = 100$ to a more smooth and connected distribution at $\alpha = 10000$.

All of these criteria turn out to be less useful for mono-age populations (\autoref{fig:alpha_single_iso}).
For such populations, the age distribution favours no smoothing at all since the true distribution is a very narrow spike.
Meanwhile, the SAMD is limited in precision by the sample size and the uncertainties on the stellar parameters.
So the method described above breaks down when the true distribution is so narrow that the uncertainties in the data limit our ability to accurately recover it.
Therefore, if the SAMD shows a single peak, $\alpha$ should be kept low so that the width of the distribution is set by the uncertainty level and sample size (this is essentially what we show in \autoref{fig:single_iso_samds} where $\alpha$ is kept constant).

Given these results, the best way to choose alpha, at least for samples with extended age-metallicity distributions, is by optimizing the recovery of the metallicity distribution.
However, this is obviously not possible when the method is applied in the one-dimensional case on age distributions.
In this case one can use the values of $-\ln L$ and $R$, e.g. by looking for a solution lying beyond the steepest part of the $-\ln L$ against $R\times\alpha$ curve.
Considering what has been described here and in Appendix \ref{sec:appA}, we recommend the following recipe for calculating the SAMD:
\begin{enumerate}
    \item Determine the required step-size for $\beta$.
    Lower it until the resulting SAMD is unchanged for a few trial values of $\alpha$.
    \item Determine the required maximum value of $\beta$.
    Increase it until the resulting SAMD no longer changes significantly for a few trial values of $\alpha$.
    \item Choose a wide range of $\alpha$ values and calculate the SAMD for each one.
    \item Choose the optimal value of $\alpha$ based on, e.g., the agreement between the true and recovered metallicity distributions, by looking at the increase in $-\ln L$ compared to $\alpha = 0$, or/and by looking for the steepest part of the $-\ln L$ against $R\times\alpha$ curve.
\end{enumerate}
This recipe has been followed in all of the tests in section \ref{sec:synth_intro}, except for the mono-age populations where a common value was chosen ($\alpha = 250$ was found to be suitable).

%%%%%%%%%%%%%%%%%%%%%%%%%%%%%%%%%%%%%%%%%%%%%%%%%%

% Don't change these lines
\bsp	% typesetting comment
\label{lastpage}
\end{document}